\documentclass[3p,12 pt]{elsarticle}

\usepackage{geometry}
\usepackage{setspace}
\usepackage{mathtools}
\usepackage{amsmath}
\usepackage{amsfonts}
\usepackage{amssymb}
\usepackage{hyperref}
\usepackage{graphicx}
\usepackage{times}
\usepackage{subcaption}
\usepackage{natbib}
\usepackage{xcolor}
\usepackage{amsthm}
\theoremstyle{definition}

\newtheorem{post}{Postulate}
\usepackage[ruled,vlined]{algorithm2e}
\hypersetup{colorlinks=true,linkcolor=blue,linkbordercolor=white,citecolor=blue}

\newcommand*{\defeq}{\mathrel{\vcenter{\baselineskip0.5ex \lineskiplimit0pt
			\hbox{\scriptsize.}\hbox{\scriptsize.}}}%
	=}

\newcommand{\norm}[1]{\left\lVert#1\right\rVert}
\journal{}
\begin{document}

	\begin{frontmatter}
	
	\title{On the extension of the concept of rheological connections to a finite deformation framework using multiple natural configurations}
	
	 \author[mainaddress]{Tarun Singh}
	\author[mainaddress]{Sandipan Paul\corref{correspondingauthor}}
	\ead{sandipan.paul@ce.iitr.ac.in}
	
	\address[mainaddress]{Department of Civil Engineering, Indian Institute of Technology Roorkee, Roorkee 247667, India}
 \cortext[correspondingauthor]{Corresponding author}
	
\begin{abstract}
        The constitutive behaviors of materials are often modeled using a network of different rheological elements. These rheological models are mostly developed within a one-dimensional small strain framework. One of the key impediments of extending these models to a three-dimensional finite deformation setting is to determine how the different types of connections, i.e., a series and a parallel connection, are incorporated into the material models. The primary objective of this article is to develop an appropriate strategy to address this issue. We show that both the series and the parallel connection between two rheological elements can be modeled within a multiple natural configurations framework without changing or introducing new configurations. The difference in a series and a parallel connection is manifested in the ratio of the stress powers expended during the deformations of the associated rheological elements. Finite deformation version of some well-known rheological models have been used to demonstrate the utility of the proposed theory.  
\end{abstract}
	
	
\end{frontmatter}
	
\section{Introduction}

Since the turn of the twentieth century, rheological elements and their connections have been widely used to represent material behaviors under different loading conditions. Rheological elements were primarily introduced to the mechanics community in the context of describing the behavior of fluids~\cite{bingham1929,bingham1933,reiner1929} and with the progress of studying different material behaviors, new rheological elements were introduced and they found their applications in several branches of mechanics such as viscoelasticity~\cite{christensen2012,lakes2009}, elasto-plasticity~\cite{simo2006}, viscoplasticity~\cite{peleg1983} etc. (cf.~Altenbach~(2023)~\cite{altenbach2023} for a review on the historical perspectives of rheological models). In recent years, these models have been extended to relatively new areas such as biomechanics~\cite{glass2008,reisinger2020}, behaviors of polymers and gels~\cite{he2020,johnsen2019}, shape memory effects~\cite{ghobadi2021,li2017,wong2011}, thermal effects~\cite{bouras2018,brocker2012}, to name a few.

Typically three rheological elements are used in different combinations of series and parallel connections to model different material responses. To describe the behavior of a 1-D linear elastic material, a spring is used where its constitutive behavior obeys Hooke's law. In case of materials exhibiting time-dependent responses, a dashpot is used in which the strain rate, instead of the strain, is proportional to the applied stress. For plasticity, a frictional block is used that exhibit no motion until the stress attains a certain value (yield stress) whereas the element shows a very large deformation once the yielding condition is met. Springs in series and parallel connections are used to determine the elastic behavior of a composite material system. Rheological models are possibly most popular in describing the time-dependent behavior of materials. The commonly used viscoelastic material models such as a Maxwell model consists of a spring and a dashpot in a series connection whereas a Kelvin model uses them in a parallel connection. A network of these elements have also been extensively used in viscoelasticity, e.g., a standard linear solid (SLS) model. In case of plasticity, a spring and a frictional block are used either in a series or a parallel connection whereas adding a frictional element to the network of springs and dashpots renders a viscoplastic response. In recent years, additional elements have also been defined and introduced to this network to describe different material behaviors. For example, Lion~(2000)~\cite{lion2000} introduced a new thermal element to derive constitutive models for thermoviscoplasticity. Kie{\ss}ling \textit{et al.}~(2016)~\cite{kiessling2016} extended this type of material models to a finite deformation setting and introduced a new element that accounts for changes in volume. Based on the work of He and Hu~(2020)~\cite{he2020}, Liang~\textit{et al.}~(2021) introduced a new fractal viscoelastic element to model the poroelastic behavior of rocks. These new elements were defined in order to account for the stress-strain behavior that cannot be considered within the networks of the three traditional rheological elements. 

The analysis of the network of 1-D rheological elements requires determination of an appropriate equilibrium and a compatibility equation for a particular segment of the network. The relevant equilibrium and compatibility equations, in turn, depend on the connections between two rheological elements. To be specific, the stress for the two rheological elements remain the same for a series connection whereas the individual strains for the elements must add up to the total strain of that segment of the network. For a parallel connection, the case is reversed, i.e., the strains for the individual elements are the same whereas the sum of the stresses across these elements result in the total stress on the network (see Fig.~\ref{fig:series_n_parallel}). 

Due to the versatility and wide applications of these rheological models in 1-D small-strain theory, several attempts have been made to extend this framework to a three-dimensional finite deformation setting. Lion~(1997)~\cite{lion1997} proposed a viscoplastic model for finite deformations in elastomers by considering a multiplicative decomposition of the respective deformation gradients, associated with the networks of the rheological elements. In this work, a strain decomposition of the rheological elements in series was replaced by a multiplicative decomposition of the deformation gradient whereas for the parallel connection between these networks, the corresponding total deformation gradient was assumed to be the same. Tacit in this work is the assumption that the configurations yield the same rate of dissipation for two different thermodynamic processes. Similar ideas were explored in a series of papers in the context of viscoelasticity~\cite{huber2000} and viscoplasticity~\cite{shutov2008,shutov2017}. Along the same line, Kie{\ss}ling~\textit{et al.}~\cite{kiessling2015,kiessling2016} generalized the idea of connections to a finite deformation. In these works, any series connection was viewed as a multiplicative decomposition of the deformation gradient of that network whereas for a parallel connection, same configurations were obtained by two \emph{different} motions~(see Fig.~\ref{fig:FD_Kiessling}). Moreover, it was assumed that the total work done on the body is a sum of the work done on individual rheological elements i.e., corresponding to their respective deformation gradients. With this assumption, they showed that the total stress in case of a series connection turns out to be the sum of the stresses on individual elements, while the stresses on the individual elements are the same for the parallel connection. 

A significant difficulty in extending a network of rheological elements to that of configurations subjected to a finite deformation is finding an equivalent relation for the decomposition of the stresses (driving forces). This case primarily arises when two rheological elements are in parallel connection. It is imperative that in the theory of 1-D rheological models, the strain and stress decomposition may seem to be a fundamental aspect although they merely come from the requirement of equilibrium and compatibility conditions. Thus a direct equivalence of these conditions will lead to several different issues. For example, Kie{\ss}ling~\textit{et al.}~\cite{kiessling2015,kiessling2016} derives the analogue of a parallel connection by considering the same configurations related by two different tangent maps (and hence, deformation gradients). Needless to say that these two deformation gradients will generate the same right Cauchy-Green tensor. However, it can be shown that when two deformation gradients produce the same right Cauchy-Green tensor, their corresponding motions are related through a rigid body motion~\cite{blume1989}. As mentioned before, since the idea of a stress and strain decomposition in case of a parallel and a series connection respectively, is \emph{not a fundamental idea, but derived from the equilibrium and compatibility conditions}, the idea of a rheological connection is simply a restriction imposed on the distribution of stress powers corresponding to the deformations of the rheological elements. In this paper, we address these issues and propose a possible extension of series and parallel connections for finite deformations in the light of multiple natural configurations.

The rest of the paper is organized as follows. In \S~\ref{sec:configurations}, we review the concepts of multiple natural configurations and define relevant quantities for the subsequent developments. We briefly discuss the existing works on this topic and identify their drawbacks in \S~\ref{sec:issues}. In \S~\ref{sec:small_strain}, we propose a postulate on the distribution of the stress powers between the elements based on the study on several commonly used rheological models. This postulate acts as a fundamental tenet for our proposed theory and is extended for a series and a parallel connection within a finite deformation setting in \S~\ref{sec:finite_def_series} and~\S~\ref{sec:finite_def_parallel} respectively. The implication of these postulates are discussed in details. In \S~\ref{sec:illustrations}, we demonstrate our proposed theory with the help of three rheological models, namely a standard linear solid, an elastic-perfectly plastic and, a elastic-plastic material with strain hardening. Thereafter, the proposed theory is summarized and drawn to conclusion. In this paper, we represent scalars with lowercase Roman/Greek letters, vectors with lowercase Roman letters in boldface and, second-order tensors with uppercase Roman letters in boldface. If $\mathbf{A}$ is a second-order tensor with components $A_{ij}$ in a certain coordinate system, then, its transpose has the components $A_{ji}$. The contraction between two second order tensors $\mathbf{A}$ and $\mathbf{B}$ has the components $(\mathbf{A}\boldsymbol{:}\mathbf{B})_{ij}=A_{ik}\,B_{kj}$. The norm of a tensor $\mathbf{A}$ is defined as $\norm{\mathbf{A}}=\sqrt{\mathbf{A}\boldsymbol{:}\mathbf{A}}$.  

\section{Preliminaries}

The primary objective of this work is to extend the network of rheological elements into a finite deformation setting. For this purpose, one needs to understand how the different types of connections can be incorporated into a framework of multiple natural configurations and their interrelations through various tangent maps. Before discussing the drawbacks of the existing literature and their possible remedy, we first briefly review the theory of multiple natural configurations which acts as a fundamental framework to our proposed model.

\subsection{Review of multiple natural configurations}\label{sec:configurations}

Let us consider a simply-connected body $\boldsymbol{\mathcal{B}}$ to be a differentiable manifold. The body is mapped into a Euclidean point space and its image is called a configuration. The undeformed and current configuration of the body are denoted by $\kappa_R(\boldsymbol{\mathcal{B}})$ and $\kappa_t(\boldsymbol{\mathcal{B}})$ respectively. Let $\mathbf{X}$ and $\mathbf{x}$ denote the corresponding position vectors of a particle of the body in its undeformed and current configuration, respectively. A motion is defined by the map $\mathbf{x}\defeq\boldsymbol{\mathcal{X}}(\mathbf{X},t)$. The deformation gradient, $\mathbf{F}$ takes a tangent vector from the undeformed configuration and maps it into the tangent space of  $\kappa_t(\boldsymbol{\mathcal{B}})$. The deformation gradient can be defined as
\begin{equation}\label{eq:def_grad}
    \mathbf{F}\defeq \dfrac{\partial \boldsymbol{\mathcal{X}}(\mathbf{X},t)}{\partial \mathbf{X}}.
\end{equation}
The right Cauchy-Green tensor, defined as $\mathbf{C}\defeq \mathbf{F}^{T}\,\mathbf{F}$, serves as a metric of the undeformed configuration of the body. Based on this metric, the Green strain tensor can be defined as
\begin{equation}\label{eq:Green_strain}
    \mathbf{E}\defeq \dfrac{1}{2}(\mathbf{C}-\mathbf{I})\quad \text{where $\mathbf{I}$ is a unit second-order tensor in $\mathbb{R}^3$}.
\end{equation}
It is worth noting that the undeformed and deformed configurations are both Euclidean (flat) spaces and therefore, the deformation gradient satisfies the condition $\text{curl}(\mathbf{F})=0$. The velocity gradient can be defined as
\begin{equation}\label{eq:vel_grad}
    \mathbf{L}\defeq\nabla_x\,\mathbf{v} = \boldsymbol{\dot{\mathbf{F}}}\,\mathbf{F}^{-1}\quad \text{where the velocity}~ \mathbf{v}\defeq \dfrac{D\,\mathbf{x}(\mathbf{X},t)}{Dt}\: 
\end{equation}

\begin{figure}[h]
    \centering
    \includegraphics[width=0.5\textwidth]{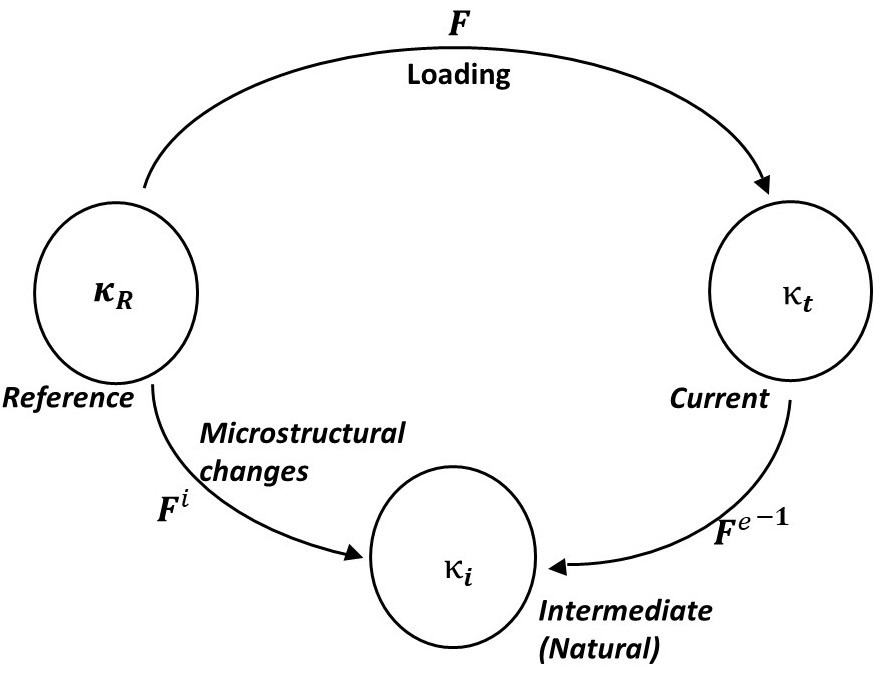}
    \caption{Multiple natural configurations and the associated tangent maps}
    \label{fig:configurations}
\end{figure}

Traditionally, to model a dissipative process, a multiplicative decomposition of the deformation gradient is used, i.e., $\mathbf{F}=\mathbf{F}^e\,\mathbf{F}^i$ where $\mathbf{F}^e$ denotes an elastic part of the deformation gradient whereas $\mathbf{F}^i$ denotes its inelastic (or, dissipative) counterpart. The elastic and inelastic parts of the deformation gradient are simply tangent maps between the respective configurations as shown in Fig.~\ref{fig:configurations}. In general, these tangent maps are not integrable and hence, it is not possible to define a deformation map between these configurations. In order to understand the physical significance of these tangent maps, we assume the existence of a family of natural configurations resulting from microstructural changes. Following Rajagopal and Srinivasa~(1998,2004)~\cite{rajagopal1998i,rajagopal1998ii,rajagopal2004}, an instantaneous elastic unloading of a particle (or an infinitesimal neighbourhood around it) in the current configuration of the body takes it into its stress-free, natural configuration, denoted by $\kappa_i(\boldsymbol{\mathcal{B}})$. It is imperative that these neighborhoods in $\kappa_i(\boldsymbol{\mathcal{B}})$ need not be globally compatible and therefore, neither $\text{curl}(\mathbf{F}^e)$ nor $\text{curl}(\mathbf{F}^i)$ needs to be zero. 
Similar to the case of total deformation, one can define the right Cauchy-Green tensor like quantities associated with $\mathbf{F}^e$ and $\mathbf{F}^i$ as
\begin{equation}\label{eq:Ce_Cp}
    \mathbf{C}^e\defeq \mathbf{F}^{e^T}\,\mathbf{F}^e; \quad \mathbf{C}^i\defeq \mathbf{F}^{i^T}\,\mathbf{F}^i
\end{equation}
and the Green strain tensor like quantities such as
\begin{equation}\label{eq:Ee_Ep}
    \mathbf{E}^e\defeq \dfrac{1}{2}(\mathbf{C}^e-\mathbf{I}) \quad \text{and,} \quad \mathbf{E}^i\defeq \dfrac{1}{2}(\mathbf{C}^i-\mathbf{I}).  
\end{equation}
The left Cauchy-Green tensor and its elastic and inelastic counterparts can be defined as
    \begin{equation}\label{eq:BeBp}
        \mathbf{B}\defeq\mathbf{F}\,\mathbf{F}^T;\quad \mathbf{B}^e\defeq\mathbf{F}^e\,\mathbf{F}^{e^T};\quad \mathbf{B}^i\defeq\mathbf{F}^i\,\mathbf{F}^{i^T}.  
    \end{equation}
In a similar manner, the elastic and inelastic velocity gradient can be defined as
\begin{equation}\label{eq:Le_Lp}
    \mathbf{L}^e = \boldsymbol{\dot{\mathbf{F}^e}}\,\mathbf{F}^{{e}^{-1}} \quad \text{and,} \quad \mathbf{L}^i = \boldsymbol{\dot{\mathbf{F}^i}}\,\mathbf{F}^{{i}^{-1}}. 
\end{equation}
These are related to the total velocity gradient through
\begin{equation}\label{eq:vel_grad_rel}
    \mathbf{L}=\mathbf{L}^e+\mathbf{F}^e\,\mathbf{L}^i\,\mathbf{F}^{e^{-1}}.
\end{equation}
 Let us decompose the velocity gradient into its symmetric and anti-symmetric parts, known as a rate of deformation tensor, $\mathbf{D}$ and a spin tensor, $\mathbf{W}$ as
    \begin{equation}\label{eq:L_D_W}
        \mathbf{D}\defeq\dfrac{1}{2}\left(\mathbf{L}+\mathbf{L}^T\right); \quad \mathbf{W}\defeq\dfrac{1}{2}\left(\mathbf{L}-\mathbf{L}^T\right).
    \end{equation}
    In the subsequent development, the rate of deformation tensor, $\mathbf{D}$ plays a key part and hence, it is necessary to define its elastic and plastic counterparts as
    \begin{equation}\label{eq:DeDp}
        \mathbf{D}^e\defeq\dfrac{1}{2}\left(\mathbf{L}^e+\mathbf{L}^{e^T}\right);\quad \mathbf{D}^i\defeq \dfrac{1}{2}\,\left(\mathbf{F}^e\,\mathbf{D}^i\,\mathbf{F}^{e^{-1}}+\mathbf{F}^{e^{-T}}\,\mathbf{L}^{i^T}\,\mathbf{F}^{e^{T}}\right) \:\text{with}\:\mathbf{D}=\mathbf{D}^e+\mathbf{D}^i.
    \end{equation}

In terms of thermodynamic processes, the elastic part of the deformation gradient is associated with a non-dissipative process whereas the inelastic deformation gradient accounts for the microstructural changes in the body leading to a dissipative process. Needless to say that every elastic unloading from the current configuration of the body will result in a different natural configuration whenever the microstructure of the body changes. The evolution of these microstructural changes is governed by an additional thermodynamic principle such as a maximum rate of dissipation criterion~\cite{rajagopal2004}. For the elastic process, we assume the existence of a Helmholtz potential function, $\psi$ that governs the elastic stress-strain response. Therefore, the total response of the body can be viewed as a family of elastic responses measured from different natural configurations. The total rate of work done on the body, thus, consists of a rate of change of stored strain energy, defined by $\dot{\psi}$ and a rate of dissipation function, $\xi$, i.e.,
\begin{equation}\label{eq:work_decomposition}
    \dot{W}=\mathbf{S}\boldsymbol{:}\boldsymbol{\dot{E}}=J\,\mathbf{T}\boldsymbol{:}{\mathbf{D}} = \,\dot{\psi} + \xi
\end{equation}
where $\mathbf{S}$ and $\mathbf{T}$ are the second Piola-Kirchhoff and Cauchy stress tensor respectively. $J$ is the Jacobian of the deformation, i.e., $J=\text{det}(\mathbf{F})$.  To determine the complete response of a body, one needs to specify the Helmholtz potential as well as the rate of dissipation function. The kinematic variable such as the elastic and inelastic strain tensors can be obtained by imposing a maximum rate of dissipation criterion.

\subsection{Issues with extending the connections of rheological models to a finite deformation theory}\label{sec:issues}

Due to the extensive use of rheological elements to model material response in a small-strain theory, several attempts have been made to extend it to finite deformation. Although the constitutive behavior of individual rheological elements and their counterparts in a finite deformation theory are well-known, the problem arises when these elements are used together in either a series or a parallel connection. In this paper, we will restrict our attention to the three traditionally used rheological elements, i.e., a spring, a dashpot and a frictional block. 

\begin{figure}[h]
    \begin{subfigure}{0.8\textheight}
		\centering
		\includegraphics[width=0.8\linewidth]{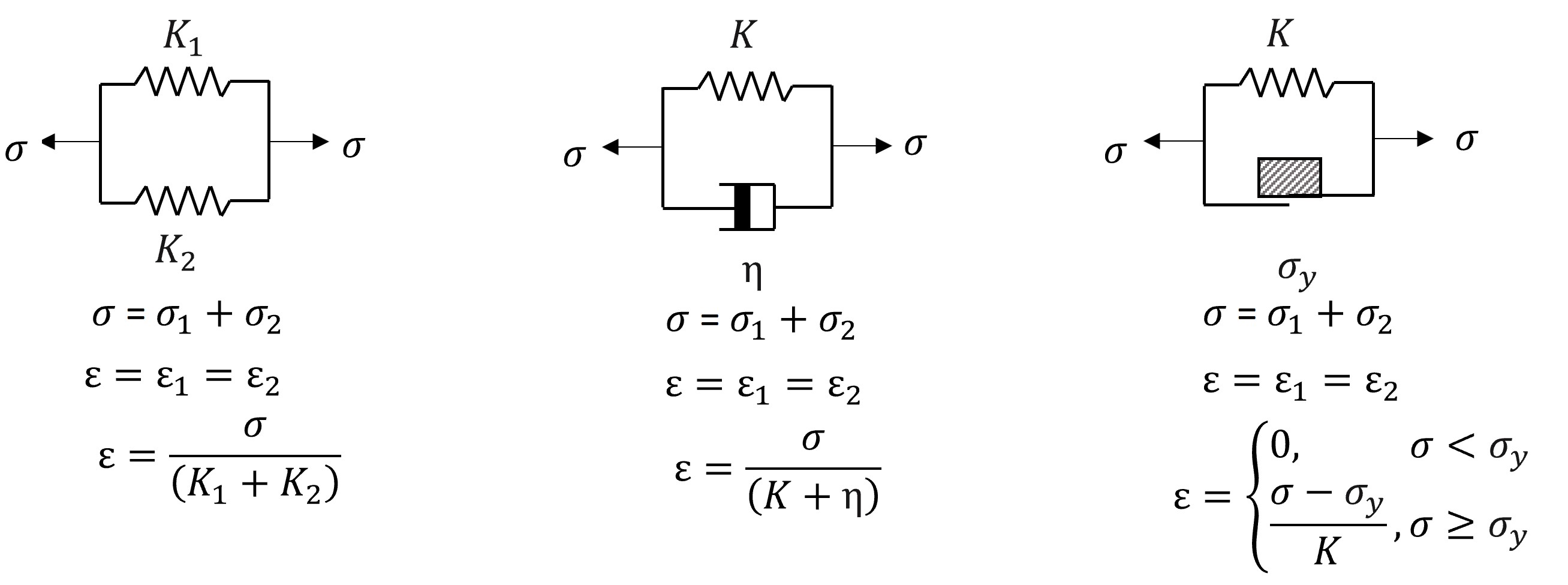}
            \subcaption{Series connection}
            \label{fig:series}
	    \end{subfigure} \\ 
		\begin{subfigure}{0.8\textheight}
		\centering
		\includegraphics[width=0.8\linewidth]{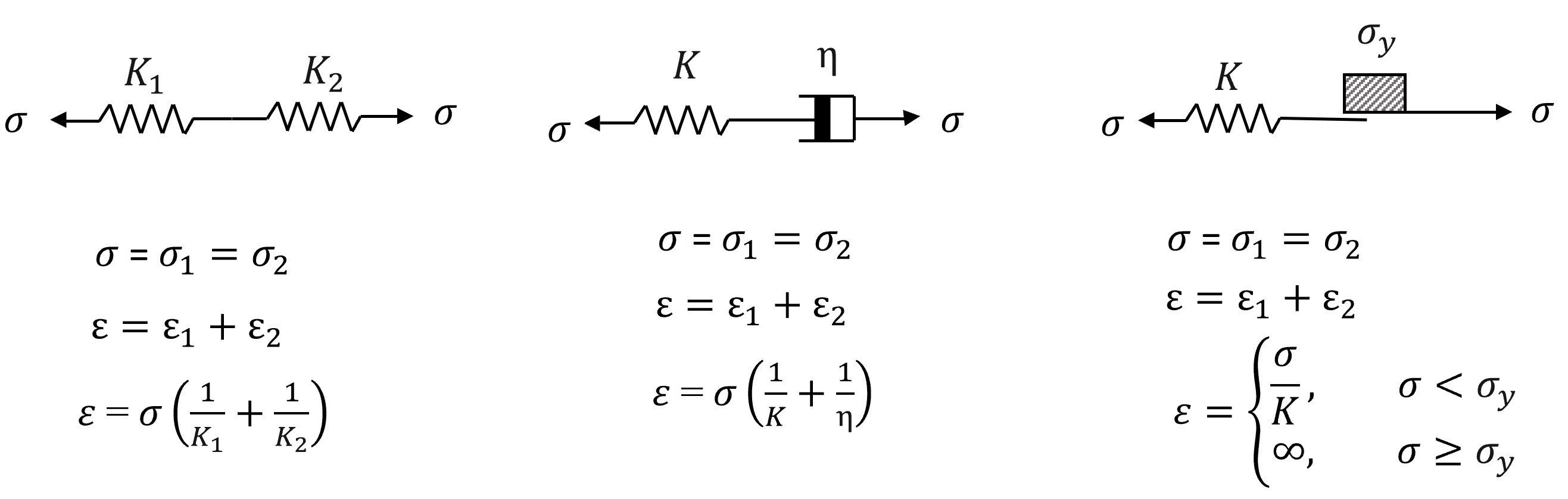}
            \subcaption{Parallel connection}
            \label{fig:parallel}
    \end{subfigure}
    \caption{Different rheological elements, viz., a spring, a dashpot and a frictional block in a series and a parallel connection along with the stress and strain distribution and, the respective governing equations.}
    \label{fig:series_n_parallel}
\end{figure}

Let us start our discussion with a series connection between the different rheological elements. In case of a series connection, the equilibrium equation implies that the stresses across the rheological elements are the same and are equal to the applied stress on the system. On the other hand, the strains across the rheological elements must add up to the total strain in order to fulfill the compatibility condition. The equilibrium equation and the compatibility condition together provide the governing equations for the system. These equations along with the constitutive behavior for different systems are summarized in Fig.~\ref{fig:series}.

\begin{figure}[h]
    \centering
    \includegraphics[width=0.9\textwidth]{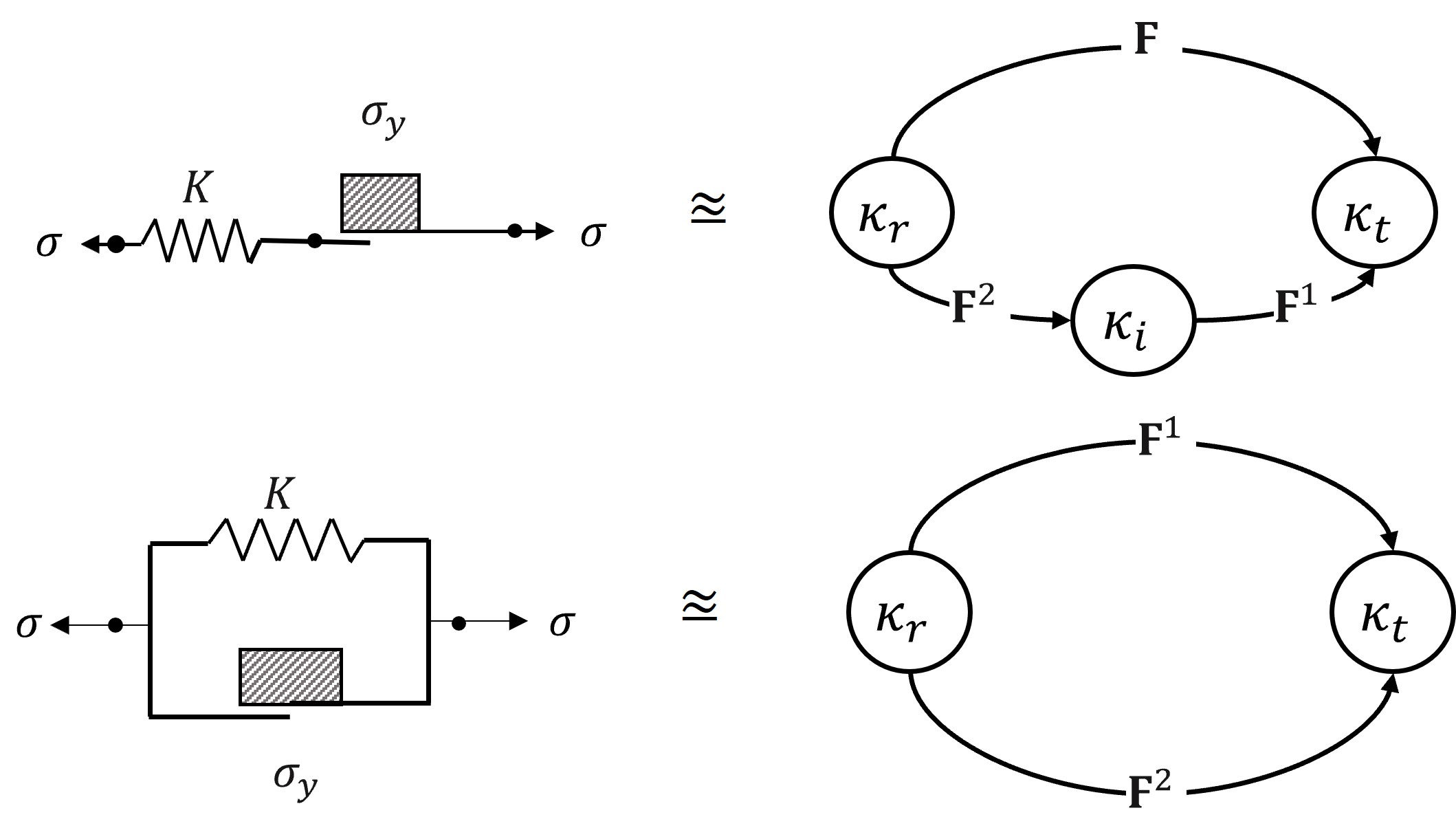}
    \caption{Traditionally used analogous configurations and their associated tangent maps to a network of rheological elements in a series and a parallel connection for large deformations.}
    \label{fig:FD_Kiessling}
\end{figure}

The most common method used in extending these material models to finite deformation (cf. Kie{\ss}ling \textit{et al.}(2015, 2016)~\cite{kiessling2015,kiessling2016}) is to consider a motion or a tangent map in place of a rheological element to take into account the strain across this element. Different configurations have been assumed to exist at the nodes between the rheological elements as shown in Fig.~\ref{fig:FD_Kiessling}. For example, in case of a series connection between a spring and a frictional block, an intermediate (natural) configuration has been considered at the node between the spring and the friction block addition to the undeformed and current configurations. A plastic deformation gradient takes the tangent vectors from the undeformed to the natural configuration whereas its elastic counterpart gradient takes the tangent vectors from the intermediate to the current configuration. Note that this particular arrangement of configurations comply with the traditional multiplicative decomposition of the deformation gradient and hence, the theory of multiple natural configurations. Needless to say that in this case the elastic and plastic strain tensors (or, rate of deformation tensors) sum up to the total strain (or rate of deformation) as is commonly derived from a multiplicative decomposition of the deformation gradient. Kie{\ss}ling \textit{et al.}~(2016) showed that if the total work done on the body is assumed to have two different components corresponding to the two rheological elements, the stresses for the two thermodynamic processes are related through only a scalar multiplies. Therefore, this system is a direct analogue of the series connection between a spring and a frictional block in case of 1-D small strain theory. 

In case of a parallel connection between two rheological elements, the same procedure has been followed. Unlike the series connection, the stresses across the rheological elements in a parallel connection add up to the total stress applied on the system whereas the total strain remains equal to the individual strains across the corresponding elements as shown in Fig.~\ref{fig:parallel}. Since a model like this contains only two nodes connected by two rheological elements, only the undeformed and current configurations have been assumed to exist in this case. Instead of an additional configuration, two different thermodynamic processes have been considered between the same two configurations with the same deformation gradient, i.e., $\mathbf{F}^1=\mathbf{F}^2$. Since the configurations are related through two different deformation gradients, they ought to generate the same amount of strain. Following a routine calculation, one can easily show that the stresses associated with the two thermodynamic processes add up to the total stress on the body whereas the strains for these two motions are the same. Therefore, this setup also complies with the traditional governing equations for an equivalent rheological system in small stain setting.

Although these two setups in a finite deformation setting appears to be correct, at least mathematically, they suffer from some fundamental fallacies. The crux of the extensions to finite deformation theory as mentioned earlier is that the configurations and the associated motions will have the same stresses across the elements in case of a series connection and same strains in case of a parallel connection. The assumption of a distribution of the total work done on the body then implies that the strains and the stresses add up to the total strain and stress on the rheological system for the respective connections. These two conditions are, however, \emph{not fundamental} in developing a material model but are derived from the respective equilibrium and compatibility conditions per the arrangement of the rheological elements. Moreover, this method suffers from a rather basic error particularly in case of a parallel connection.

In case of a parallel connection between two rheological elements, it has been assumed that the two configurations are related through two different motions. This assumption bears two primary errors:
\begin{itemize}
    \item It is well known that a dissipative process, e.g., a plastic deformation typically denotes microstructural changes and it renders the corresponding configurations incompatible except for a fully homogeneous plastic deformation. In fact, measure of this incompatibility provides the content of dislocation lines~\cite{acharya2000,gurtin2001,paulFreed2020gnd} in case of a plastic deformation. On the other hand, an elastic process is \emph{always} non-dissipative and a flat configuration undergoing an elastic process remains flat. Therefore, if two configurations are related by an elastic and a plastic deformation process, the former would leave the deformed configuration flat whereas the latter will introduce a torsion to the current configuration. Hence, such a system will not be valid in general.   

    \item This problem also persists when two non-dissipative processes are considered, i.e., in case of a parallel connection between two springs. Since the same configurations are related through two different motions, they ought to produce the same amount of strain. In fact, this has been considered as the fundamental basis for arranging the respective configurations in case of a parallel connection ( see Fig.~\ref{fig:FD_Kiessling}). Let us consider that $\mathbf{F}^1$ and $\mathbf{F}^2$ are the deformation gradients corresponding to the two thermodynamic processes. Therefore, the right Cauchy Green tensor (and, the Green strain) remain the same for these two cases and they are related through
    \begin{equation}\label{eq:C_def_grad_parallel}
        \mathbf{F}^{1^T}\,\mathbf{F}^1=\mathbf{F}^{2^T}\,\mathbf{F}^2=\mathbf{C}.
    \end{equation}
    Blume~(1989)~\cite[\S 2, Th. 2.1]{blume1989} has shown that the two deformation gradients produce the same right Cauchy-Green tensor if and only if they are related through a rigid body motion, i.e.,
    \begin{equation}
     \boldsymbol{\mathcal{X}^1 (\mathbf{X},t)}= \mathbf{Q}\, \boldsymbol{\mathcal{X}^2(\mathbf{X},t)} + \mathbf{d}\quad\text{where}\: \mathbf{Q}\,\boldsymbol{\epsilon} \: \boldsymbol{\mathcal{O}}^+.
    \end{equation}
    Here $\boldsymbol{\mathcal{X}^1 (\mathbf{X},t)}$ and $ \boldsymbol{\mathcal{X}^1 (\mathbf{X},t)}$ are the motions corresponding to the deformation gradients $\mathbf{F}^1$ and $\mathbf{F}^2$, respectively and the vector $\mathbf{d}$ represents a rigid body translation. Moreover, in  Kie{\ss}ling \textit{et al.}~(2016) the rotation tensor has $\mathbf{Q}$ been assumed to be an identity tensor. Since the two processes eventually refer to the motions related by only a rigid body motion, this extension of a parallel connection to finite deformation theory is also not valid when two springs are under consideration.
\end{itemize}

Although in case of a series connection, the methodology of Kie{\ss}ling \textit{et al.}~(2015, 2016) seem to work, the issues are very evident from the above discussion when the rheological elements are in a parallel connection concept. Therefore, the aim of the present work is to develop a sound theory to extend the concept of rheological connections that satisfy all the underlying assumptions of a finite deformation theory.

\section{Different connections between rheological elements within a small-strain theory}\label{sec:small_strain}

To model the response of a body undergoing finite deformation, the framework of multiple natural configurations deem to be very suitable due to the physical significance of the relevant configurations. This framework is particularly useful when the deformation involves a dissipative process resulting from microstructural changes. As discussed earlier, the existing models needed to be altered as per the arrangement of the rheological elements. On the other hand, an advantage of using this framework is that the arrangement of the configurations and the associated tangent maps remain the same irrespective of the nature of the rheological elements involved or the connection between them. Moreover, the previously proposed models were developed keeping the stress and strain distribution across the elements in mind. The stress and strain distribution, however, are merely derived from the equilibrium and compatibility conditions. Therefore, this type of models lead to some fundamental discrepancies as discussed earlier. 

It is well known that any deformation process is governed by the amount of energy (or, stress power) available to the associated thermodynamic process and how much of this energy is stored as strain energy or gets dissipated to accommodate the microstructural changes. Based on the total work done, energy dissipated during a deformation process and the strain energy stored in the body, the strains and the stresses corresponding to a particular deformation are determined. Keeping this in mind, we propose the following postulate for the rheological models which will be extended to a finite deformation theory in \S~\ref{sec:finite_def}.
\begin{post}\label{Post:0}
    For a connection of 1-D rheological elements within a small-strain theory, the stress power expended during the associated thermodynamic processes is distributed according to the ratio of strain rates for a series connection and the ratio of stresses across the elements for a parallel connection respectively. If the strain rates across the two rheological elements are denoted by $\dot{\epsilon}_1$ and $\dot{\epsilon}_2$ with the corresponding stresses $\sigma_1$ and $\sigma_2$, then the ratio of stress power expended during the deformation of these elements are given by   
    $$
        {\dot{W}_1/\dot{W_2}}=\begin{cases}
			\big|\dot{\epsilon}_1/\dot{\epsilon}_2\big|, & \text{for a series connection}\\
                \big|\sigma_1/\sigma_2\big|, & \text{for a parallel connection}.
		 \end{cases}
        $$
\end{post}
Here \big|$\cdot$\big| represents magnitude of a scalar variable.

\subsection{Commonly used rheological models in the context of the proposed postulate}
By post.~\ref{Post:0}, the total stress power equals to the sum of the individual rates of stored energy or dissipation by each rheological element, i.e.,
 \begin{align}\label{eq:power_total}
      \dot{W}&=\dot{W}_1+\dot{W}_2 \tag{11a} \\
      &= \sigma_1\cdot\dot{\epsilon}_1+\sigma_2\cdot\dot{\epsilon}_2.\tag{11b}
 \end{align}

\subsubsection{Connections between two linear springs}
Since springs represent elastic (non-dissipative) processes, the ratio $\dot{W}_1/\dot{W}_2$ signifies the ratio of the rate of stored energy associated with the individual springs.

    \begin{itemize}
    \item Series connection\\
         Here the rate of stored energy associated with the two springs are given as 
         \begin{align}\label{eq:spring_spring_series_power}
             \dot{W}_1&={\sigma}_1\cdot\dot{{\epsilon}}_1\tag{12a}\\
             \dot{W}_2&={\sigma}_2\cdot\dot{{\epsilon}}_2\tag{12b}.
         \end{align}
          \setcounter{equation}{12}
         Now according to post.~\ref{Post:0}, these rates of stored energy must be distributed as per the ratio of strain rates across them, therefore, $\dot{W}_1/\dot{W}_2=\big|\dot{{\epsilon}}_1/\dot{{\epsilon}}_2$\big|. From eqn.~\eqref{eq:spring_spring_series_power}, it is evident that the ratio of stresses must be $1$ for the condition prescribed by post.~\ref{Post:0} to be satisfied, i.e., $\sigma_1=\sigma_2$. Note that \emph{both} the magnitude and sign of the stresses will be the same since the sign of stress and strain across an element must be the same.
         
     \item Parallel connection \\
        The only difference in case of a parallel connection between two springs is that the total rate of work done is distributed according to the stress ratio, instead of a ratio of strain rates. Therefore, in this case, we have
        \begin{align}
        \dot{W}_1/\dot{W}_2={\sigma}_1\cdot\dot{{\epsilon}}_1/{\sigma}_2\cdot\dot{{\epsilon}}_2 \implies \dot{\epsilon}_1/\dot{\epsilon}_2=1, \quad \text{since}\: \dot{W}_1/\dot{W}_2=\big|{\sigma}_1/{\sigma}_2\big|.
        \end{align}
    \end{itemize}

\subsubsection{Connections between a linear spring and a dashpot} 
    Since a spring-dashpot system involves both a non-dissipative and a dissipative process for the spring and the dashpot respectively, the distribution of total power has different physical significance. In this system, a portion of the total power is used to increase the strain energy in the system (spring) and the rest is dissipated in the form of heat (dashpot). We will now see how different connections affect this distribution. Here, $\dot{W}_1=\dot{\psi}$ and $\dot{W}_2=\xi$.
     \begin{itemize}
     \item Series connection \\
     For a series connection between the rheological elements,
          \begin{equation}\label{eq:spring_dashpot_series}
          \dot{W}_1/\dot{W}_2=  \dot{\psi} / \xi=\big|\dot{{\epsilon}}_1/\dot{{\epsilon}}_2\big|
         \end{equation}
    Now using the definition of $\dot{W}_1$ and $\dot{W}_2$ and Eqn.~\eqref{eq:spring_dashpot_series}, one can write
            \begin{align}
                    \dot{W}_1/\dot{W}_2={\sigma_1}\cdot\dot{{\epsilon}}_1/{\sigma}_2\cdot\dot{{\epsilon}}_2 \,\implies\: {\sigma}_1 ={\sigma}_2={\sigma}
            \end{align}

    \item Parallel connection \\
         In a similar manner, since $\dot{W}_1/\dot{W}_2=\dot{\psi} / \xi=\sigma_1/\sigma_2$
         in case of a parallel connection between the said elements, one can easily show that 
         \begin{align}
            \dot{W}_1/\dot{W}_2={\sigma}_1\cdot\dot{{\epsilon}}_1/{\sigma}_2\cdot\dot{\mathbf{\epsilon}}_2 \,\implies\: \dot{\epsilon_1} = \dot{\epsilon}_2=\dot{\epsilon}.
            \end{align}
    \end{itemize}
    
\subsubsection{Connections between a linear spring and a frictional block}
        Although the connection between a linear spring and a frictional block is similar in nature in the sense that they both involve a dissipative (dashpot/frictional block) and a non-dissipative (spring) process, physically the frictional block works in a very different way. This is a special case since for a frictional block, the strain shows a discontinuity such as
        $$
        {\epsilon}=\begin{cases}
			0, & {{\sigma}<{\sigma}_y}\\
            \infty, & {{\sigma}\ge{\sigma}_y}
		 \end{cases}
        $$
        We show that even though the underlying mathematics is a bit different, the post.~\ref{Post:0} still holds in this case. 
        
        \begin{itemize}
                \item Series connection \\
                 Here $\dot{W}_1=\dot{\psi}={\sigma}_1\cdot\dot{{\epsilon}}_1$ and $\dot{W}_2=\xi={\sigma}_2\cdot\dot{{\epsilon}}_2.$ Now for a series connection  since $\dot{W}_1/\dot{W}_2=\dot{\epsilon}_1/\dot{\epsilon}_2$, we have  
                \begin{align} \label{eq:spring_slider_series}
                {\dot{W}_1/\dot{W}_2}=\begin{cases}
    		\infty, & {{\sigma}<{\sigma}_y}\\
                $0$, & {{\sigma}\ge{\sigma}_y.}
		      \end{cases}
                \end{align}
        Eqn.~\eqref{eq:spring_slider_series} implies that when $\sigma<\sigma_y$, the entire power is used in the elastic deformation process whereas when the total stress is greater than the yield stress of the frictional block, the block slides and the entire power is expended on the dissipative, plastic deformation process.  

    \item Parallel connection \\
        In case of a parallel connection, the total power is distributed according to the stresses across the rheological elements, i.e., $\dot{W}_1/\dot{W}_2={\sigma_1}/{\sigma_2}$. When $\sigma<\sigma_y$, the entire power is taken up by the frictional block. When the requirement of the frictional block is satisfied to its maximum capacity (yield stress), the rest is stored as a strain energy. Therefore, in this case, we have
            \begin{align}  \label{eq:spring_slider_parallel}
            {\dot{W}_1/\dot{W}_2}=\begin{cases}
		\infty, & {{\sigma}<{\sigma}_y}\\
            \alpha & {{\sigma}\ge{\sigma}_y}\; \text{where}\; \alpha,\:\epsilon\:\mathbb{R^+}.
		 \end{cases}
            \end{align}

    \end{itemize}
    Hence, post.~\ref{Post:0} holds for all the traditional rheological models. Motivated by this, we intend to extend this framework to a finite deformation theory. 

    \section{Rheological connections in a finite deformation framework}\label{sec:finite_def}

    In this section, we elucidate the implication of post.~\ref{Post:0} in the context of finite deformation. As mentioned in \S~\ref{sec:configurations}, the framework of multiple natural configurations is very suitable in describing the response of a body and evolution of its microstructures in case of a finite deformation. Before going into the extension of material models consisting of rheological elements into a finite deformation theory, let us first talk about the relevant constitutive equations, their derivations and the evolution equations that govern a thermodynamic process. In this section, we do not impose any restriction on the type of thermodynamic processes, i.e., $\mathbf{F}^1$ and $\mathbf{F}^2$, can be associated with any thermodynamic process irrespective of their dissipative or non-dissipative nature. Whenever a combination of a dissipative and a non-dissipative process is to be considered, we shall choose $\mathbf{F}^e$ to represent the deformation gradient associated with a non-dissipative process.

    \subsection{General constitutive and evolution equations for multiple natural configurations}\label{sec:general_constitutive}

    To understand the thermodynamic processes, we first need to specify constitutive assumptions for the Helmholtz potential function, $\psi$ and the rate of dissipation function, $\xi$. Needless to say that these two quantities are related through eqn.~\eqref{eq:work_decomposition} that needs to be satisfied at all times. Moreover, we consider the rate of dissipation function to be admissible only when it is non-negative. 

    Since the Helmholtz potential function, $\psi$ describes the response associated with the motion denoted by $\mathbf{F}^e$ measured from a particular natural configuration, it must be dependent on $\mathbf{F}^e$ and $\mathbf{F}^i$. This dependence, however, does not require to be an explicit one and therefore, one can use any one of the following arguments for $\psi$: $(\mathbf{F},\mathbf{F}^i)$, $(\mathbf{F}^e,\mathbf{F}^i)$ or, $\mathbf{F}^e$. In this derivation, we choose the first argument and modify it to represent the strain-like quantities defined in eqn.~\eqref{eq:Ee_Ep}. Hence, the constitutive equation for the Helmholtz potential function can be written as
    \begin{equation}\label{eq:psi_constitutive}
        \psi=\bar{\psi}\,(\mathbf{E}, \mathbf{E}^i).
    \end{equation}
    Since the rate of dissipation function is associated with the evolution of the natural configurations, its constitutive relation can be specified as
    \begin{equation}\label{eq:xi_constitutive}
        \xi=\bar{\xi}\,(\mathbf{E}^i, \mathbf{D}^i)\quad \text{with}\quad \bar{\xi}\,(\mathbf{E}^i,\mathbf{0})=0.
    \end{equation}
    We assume that the elastic response of the body is that of a Green elastic solid and therefore, the second Piola-Kirchhoff stress can be written as
    \begin{equation}\label{eq:Green_elastic}
        \mathbf{S}=\rho_0\,\dfrac{\partial \bar{\psi}}{\partial \mathbf{E}}.
    \end{equation}
    Substituting eqns.~\eqref{eq:psi_constitutive}, \eqref{eq:xi_constitutive}, \eqref{eq:Green_elastic} into eqn.~\eqref{eq:work_decomposition}, we get
    \begin{equation}\label{eq:xi_constraint}
        \xi=\rho_0\,\dfrac{\partial \bar{\psi}}{\partial \mathbf{E}^i}\boldsymbol{:}\boldsymbol{\dot{E}}^i.
    \end{equation}
    To derive the evolution equation for $\boldsymbol{\dot{E}}^i$ (or, $\mathbf{D}^i$), we need to stipulate a maximum rate of dissipation criterion in addition to the laws of thermodynamics. Following Rajagopal and Srinivasa~(2004)~\cite{rajagopal2004}, this criterion requires the maximization of $\xi$ with eqn.~\eqref{eq:xi_constraint} as a constraint. By using a Lagrange multiplier method, the maximization of $\xi$ yields
    \begin{equation}\label{eq:implicit_evolution}
        \dfrac{\partial \bar{\xi}}{\partial \boldsymbol{\dot{E}^i}} = \dfrac{\lambda\,\rho_0}{1+\lambda}\,\dfrac{\partial\bar{\psi}}{\partial \mathbf{E}^i}
    \end{equation}
    where $\lambda$ is a Lagrange multiplier which can be obtained by the satisfaction of eqn.~\eqref{eq:xi_constraint} from
    \begin{equation}\label{eq:lambda}
        \dfrac{\lambda}{1+\lambda}=\dfrac{\dfrac{\partial \xi}{\partial \mathbf{E}^i}\boldsymbol{:}\boldsymbol{\dot{E}}^i}{\xi}.
    \end{equation}
    Note that eqn.~\eqref{eq:implicit_evolution} is an implicit equation in $\boldsymbol{\dot{E}}^i$. An explicit equation can be derived when the constitutive specifications are provided in a stress/driving force space. We shall discuss that case later.

    To develop analogues for the rheological models in a finite deformation theory, we first need to propose a corresponding version of post.~\ref{Post:0}. It is important to note that the stress power can be expressed as an inner product of the conjugate stress and strain rate pairs. Here we choose the Cauchy stress tensor, $\mathbf{T}$ and the deformation rate tensor $\mathbf{D}$ to write the total stress power as
    \begin{equation}\label{eq:total_stress_power_constitutive}
        \dot{W}=J\,\mathbf{T}\boldsymbol{:}\mathbf{D}.
    \end{equation}
    If other conjugate stress/strain pairs are used, the postulate needs to be altered accordingly.

 \subsection{Finite deformation model for a series connection}\label{sec:finite_def_series}

    To derive a finite deformation version of the rheological connections, we first need to determine an appropriate postulate equivalent to posts.~\ref{Post:0}. Note that the rheological models are restricted to one-dimension only and thus, the power distribution is dependent on the respective ratio of strain rates or stresses which are essentially a numerical value. In a three-dimensional deformation, however, one must take the ratio of some \textit{``magnitudes"} of these quantities to determine the distribution of the power expended. Here we use a simple $\mathcal{L}^2-$ norm of the relevant quantities; however, other definitions of the \textit{magnitudes} of these second-order tensors and hence, variations of postulates~\ref{Post:1} and \ref{Post:2} are also possible. Let us define a norm of any second-order tensor $\mathbf{A}$ as $\norm{\mathbf{A}}\defeq \sqrt{\mathbf{A}\boldsymbol{:}\mathbf{A}}$.
    
    \begin{post}\label{Post:1}
    For a series connection between two rheological elements, the stress power expended during the corresponding thermodynamic processes is distributed according to the ratio of the norm of associated rate of deformation tensors. In other words, if the motions corresponding to two rheological elements are subjected to rate of deformation tensors $\mathbf{D}_1$ and $\mathbf{D}_2$, then, the ratio of the stress power expended corresponding to the associated deformation processes is $\dot{W}_1/\dot{W}_2=\norm{\mathbf{D}_1}/\norm{\mathbf{D}_2}$. 
    \end{post}

    We assume that the post.~\ref{Post:1} holds true for \emph{any thermodynamic processes} that the elements may undergo. Therefore, for a spring, $\dot{W}_i,\: i=1 \:\text{or}\:2$, becomes the rate of change of Helmholtz potential fuction $\dot{\psi}$; on the other hand, in case of a frictional block or a dashpot $\dot{W}_i$ is the rate of dissipation, $\xi$.  Now if the ratio of $\dot{W}_i$'s are known, the energy expended for a deformation corresponding to a particular element is also known by using eqn.~\eqref{eq:work_decomposition} and therefore, the stresses and rate of deformation tensors across them can easily be obtained using the relevant laws of thermodynamics. The ratio, $\omega^s\defeq\dot{W}_1/\dot{W}_2=\norm{\mathbf{D}^e}/\norm{\mathbf{D}^i}$, however, is not known \textit{a priori} and depends on the ratio of the rate of deformation tensors. Hence, the ratio of power expended, $\omega^s$ and the corresponding deformation rate tensors need to be determined simultaneously.  

    Here we follow the same procedure as laid out in \S~\ref{sec:general_constitutive} with some minor changes. To implement post.~\ref{Post:1}, we use the rate of deformation tensors $\mathbf{D}^i$ in place of the strain rate, $\boldsymbol{\dot{E}}_i$. Since these two kinematic quantities are related through $\boldsymbol{\dot{E}^i}=\mathbf{F}^{i^T}\,\mathbf{D}^i\,\mathbf{F}^i$, they essentially bear the same information. Another major change in this derivation is its starting point. Since the ratio of stress power, $\omega^s$ is known following post.~\ref{Post:1}, one can write the rate of dissipation function in terms of the total stress power as
    \begin{equation}\label{eq:xi_constraint_series}
        \xi=\dfrac{1}{1+{\omega}^s}\,J\,\mathbf{T}\boldsymbol{:}\mathbf{D}.
    \end{equation}
    Note that the eqn.~\eqref{eq:xi_constraint_series} acts as a constraint for the rate of dissipation function $\xi$ unlike the general constraint equation~\eqref{eq:xi_constraint}. Now to find out the relevant kinematic quantities such as the rate of deformation tensor $\mathbf{D}^i$, a maximum rate of dissipation criterion needs to be implemented. Using a Lagrange multiplier method with eqn.~\eqref{eq:xi_constraint_series} as a constraint, one can write the Lagrangian as
    \begin{equation}\label{eq:lagrangian_series}
        \mathbb{L}^s=\xi+\lambda\left[\xi-\dfrac{1}{1+{\omega}^s}\left(J\,\mathbf{T}\boldsymbol{:}\mathbf{D}\right)\right]
    \end{equation}
    which needs to be maximized with respect to admissible $\mathbf{D}^i$. Note that here the stress power as well as the term ${\omega}^s$ are both functions of $\mathbf{D}^i$. A routine calculation {(see App.~\ref{sec:evolution_series})} produces the evolution equation for $\mathbf{D}^i$ in its implicit form as
    \begin{equation}\label{eq:evolution_series}
        \dfrac{\partial \bar{\xi}}{\partial \mathbf{D}^i}=\dfrac{\lambda}{1+\lambda}\left[\dfrac{{\omega}^s}{(1+{\omega^s})^2}\,\dfrac{\mathbf{D}^i}{\norm{\mathbf{D}^i}^2}\,\left(J\,\mathbf{T}\boldsymbol{:}\mathbf{D}\right)+\dfrac{1}{1+{\omega}^s}\left(J\mathbf{T}\right)\right].
    \end{equation}
    The Lagrange multiplier can be found by satisfaction of the constraint equation and has the same form as in eqn.~\eqref{eq:lambda}. The right hand side of eqn.~\eqref{eq:evolution_series} is essentially a driving force, denoted by $\mathbf{T}^i$, for the motion associated with the rate of deformation tensor $\mathbf{D}^i$. {A detailed discussion on this driving force is provided in \S~\ref{sec:EM}. 

    \subsection{Discussion on the driving force and its relation with the Eshelby energy-momentum tensor}\label{sec:EM}

    Let us understand the implications of each of the thermodynamic processes involved in this framework. As mentioned earlier, in case of a multiple natural configurations, one can imagine two thermodynamic processes leading to a total deformation with a Cauchy stress tensor $\mathbf{T}$ and a deformation gradient $\mathbf{F}$. Although a multiplicative decomposition of the deformation gradient gives us the associated kinematic quantities for the individual thermodynamic processes generating stress powers $\dot{\psi}$ and $\xi$ respectively, their kinetic conjugates are still not understood. Since the dissipative process involves microstructural changes and plays an important role in causing several macroscopic changes such as plastic deformation, damage, fracture etc., the corresponding kinetic quantity has garnered much attention. In context of an elliptical inclusion, this \emph{driving force} behind the dissipative process was first addressed by Eshelby~\cite{eshelby1951,eshelby1956,eshelby1975} and then it was extended to other mechanical theories as well~\cite{epsteinmaugin1990,epstein1996,cleja2000,cermelli2001}. The driving force can be written as~\cite{gupta2007}
    \begin{equation}\label{eq:Eshelby_traditional}
        {\boldsymbol{\mathcal{E}}\defeq \psi\,\mathbf{I}-\mathbf{F}^T\,\mathbf{P}},
    \end{equation}
    where $\mathbf{P}$ is the first Piola-Kirchhoff stress. $\boldsymbol{\mathcal{E}}$ is often termed as an Eshelby energy-momentum tensor and can be shown to be equivalent to a \emph{configurational force}.

    In our case, the right hand side of eqn.~\eqref{eq:evolution_series} acts as a driving force for the motion associated with $\mathbf{D}^i$. A similar expression for the driving force was derived by Rajagopal and Srinivasa~(1998)~\cite[eqn.~(19)]{rajagopal1998ii} for a multiple natural configurations framework in the context of the theory of plasticity. One can write the driving force governing the dissipative process as
    \begin{equation}\label{eq:Ti_def}
        \mathbf{T}^i\defeq \dfrac{1}{J^i}\,\dfrac{\partial \bar{\xi}}{\partial \mathbf{D}^i}\quad \text{with}\quad J^i\,\mathbf{T}^i\boldsymbol{:}\mathbf{D}^i=\xi
    \end{equation}
    where $J^i=\text{det}(\mathbf{F}^i)$ with $J=J^e\,J^i$. Now from eqn.~\eqref{eq:evolution_series} and using post.~\ref{Post:1}, this driving force can be expressed as
    \begin{equation}\label{eq:Ti_final}
        \mathbf{T}^i=\dfrac{1}{J^i\,(1+\omega^s)}\left[\dot{\psi}\dfrac{\mathbf{D}^i}{\norm{\mathbf{D}^i}^2}\,+J\mathbf{T}\right].
    \end{equation}
    From eqns.~\eqref{eq:Eshelby_traditional} and \eqref{eq:Ti_final}, one can observe that although the traditionally defined Eshelby energy-momentum tensor is not the same as our driving force function, $\mathbf{T}^i$, there are some discernible similarities between these two. Both the Eshelby energy-momentum tensor $\boldsymbol{\mathcal{E}}$ and the driving force $\mathbf{T}^i$ contain a Cauchy stress term or another version of it, pulled back into the natural configuration. On the other hand, $\boldsymbol{\mathcal{E}}$ contains a term involving the Helmholtz potential function while the driving force has its time-derivative in its expression. This is due to the fact that a specific form for $\psi$ was assumed in the derivation of the Eshelby energy-momentum tensor whereas we derive the driving force function from the expression of a \emph{rate} of dissipation function. In their discussion of the form of the Eshelby energy-momentum tensor, Rajagopal and Srinivasa~(2005)~\cite{rajagopal2005} discussed the same issue and provided a detailed explanation. Another key difference between the Eshelby energy-momentum tensor, $\boldsymbol{\mathcal{E}}$ and the driving force, $\mathbf{T}^i$ is the co-efficient multiplied with the Helmholtz potential function (or its rate). In case of our driving force function $\mathbf{T}^i$, a normalized form of the rate of deformation tensor $\mathbf{D}^i$ is multiplied with $\dot{\psi}$ unlike a unit second-order tensor used in the expression of $\boldsymbol{\mathcal{E}}$. The term $\mathbf{D}^i/\norm{\mathbf{D}^i}$ represents the \emph{``direction"} of $\mathbf{D}^i$ and plays a significant role in the subsequent derivation. 

    Now let us look at the consequence of eqn.~\eqref{eq:Ti_def} for the driving force associated with the motion involving $\mathbf{D}^i$. Contracting eqn.~\eqref{eq:evolution_series} with $\mathbf{D}^i$ and using eqn.~\eqref{eq:Ti_def}, we arrive at
    \begin{equation}\label{eq:stress_relations_series}
        \mathbf{T}=\left(\dfrac{J^i}{J}\right)\mathbf{T}^i
    \end{equation}
    where $\mathbf{T}^i$ is the driving stress behind the inelastic process, resulting in $J^i\,\mathbf{T}^i\boldsymbol{:}\mathbf{D}^i={\xi}$. For detailed calculations, see App.~\ref{sec:evolution_series}.

    Although the result~\eqref{eq:stress_relations_series} is quite similar to the models discussed in \S~\ref{sec:issues}, our theory is developed entirely within a well established framework of multiple natural configurations. The result~\eqref{eq:stress_relations_series} is a mere consequence of the post.~\ref{Post:1} and it does not depend on the arrangement of configurations as is the case in other models. Therefore, the theoretical issues appearing in other similar models are successfully averted in our theory. The utility of this framework is even more prominent when a parallel connection between the elements is considered. It is also worth noting that the series connection in the models described in \S~\ref{sec:issues} also use a multiplicative decomposition of the deformation gradient and Hence, the similarity in the results for a series connection does not come as a surprise. It is the parallel connection between the rheological elements where a completely different approach must be adopted.

    \subsection{Finite deformation model for a parallel connection between rheological elements}\label{sec:finite_def_parallel}

    As seen in the 1-D version of the rheological models in post.~\ref{Post:0}, the stress power is distributed according to the magnitude of corresponding stresses when a parallel connection between rheological elements is considered. Hence, a three dimensional, finite deformation version of this postulate must be proposed in such a way that the stress power is distributed as per the ratio of some kinetic quantity (or flux) corresponding to the two individual motions. From \S~\ref{sec:EM}, we have seen that the driving force functions, $\mathbf{T}^e$ and $\mathbf{T}^i$ govern the processes in the sense that when contracted with their kinematic counterparts, i.e., $\mathbf{D}^e$ and $\mathbf{D}^i$ respectively, the driving force functions generate the stress powers, $\dot{\psi}$ and $\xi$. In the previous case, the driving force function was obtained by maximizing the rate of dissipation function with respect to the rate of deformation tensor $\mathbf{D}^i$. Since the condition $\xi=J^i\,\mathbf{T}^i\boldsymbol{:}\mathbf{D}^i$ must always be satisfied, one can implement the criterion of a maximum rate of dissipation function by carrying out the maximization either with respect to $\mathbf{D}^i$ keeping the driving force $\mathbf{T}^i$ constant or vice versa. In case of a parallel connection we shall adopt the latter: We assume the constitutive relation for the rate of dissipation function as
    \begin{equation}\label{eq:xi_parallel}
        \xi=\hat{\xi}(\mathbf{E}^i, \mathbf{T}^i).
    \end{equation}
    Since a parallel connection between rheological elements are governed by the ratio of corresponding driving forces, we need to carry out the same process within the driving force space in this case. Now let us propose the finite deformation version of post.~\ref{Post:0} for parallel connections.

    \begin{post}\label{Post:2}
        For a parallel connection between two rheological elements, the total stress power is distributed according to the norm of the driving forces corresponding to the individual motions. If $\dot{W}_1$ and $\dot{W}_2$ represent the stress power expended during the motions associated with the respective rheological elements and, $\mathbf{T}_1$ and $\mathbf{T}_2$ denote the corresponding driving forces, then, $\dot{W}_1/\dot{W}_2=\norm{\mathbf{T}_1}/\norm{\mathbf{T}_2}$.
    \end{post}
    
    Now let us look at the implication of this postulate for a general thermodynamic process. Since except for the case of connection between two linear springs, every other connections involve a dissipative and a non-dissipative process, here we assume that $\dot{W}_2$ is to a rate of dissipation function while $\dot{W}_1$ signifies a rate of change of Helmholtz potential function. The notation for the corresponding kinematic and kinetic quantities must be altered accordingly. Let us denote the ratio of stress powers by $\omega^p$ and thus by post.~\ref{Post:2},
    \begin{equation}\label{eq:omega_parallel}
        {\omega}^p\defeq\dfrac{\dot{W}_1}{\dot{W}_2}=\dfrac{\dot{\psi}}{\xi}=\dfrac{\norm{\mathbf{T}^e}}{\norm{\mathbf{T}^i}}.
    \end{equation}
    If the total stress power is denoted by $\dot{W}$ and it follows eqn.~\eqref{eq:work_decomposition}, then eqn.~\eqref{eq:omega_parallel} implies that
    \begin{equation}\label{eq:xi_parallel}
        \xi=\dfrac{1}{(1+{\omega}^p)}\,J\,\mathbf{T}\boldsymbol{:}\mathbf{D}.
    \end{equation}
    Needless to say that eqn.~\eqref{eq:xi_parallel} acts as a constraint for the rate of dissipation function. To implement a criterion of maximum rate of dissipation, we now need to maximize $\xi$ using a Lagrange multiplier method, but this time with respect to the driving force function, $\mathbf{T}^i$. The optimizing variable is the key difference between the two frameworks corresponding to a series and a parallel connection. The Lagrangian, in this case, has the same form as before, i.e.,
    \begin{equation}\label{eq:Lagrangian_parallel}
        \mathbb{L}^p=\xi+\lambda\left[\xi-\dfrac{1}{(1+{\omega}^p)}\,J\,\mathbf{T}\boldsymbol{:}\mathbf{D}\right]
    \end{equation}
    where $\lambda$ is a Lagrange multiplier. Now maximizing the Lagrangian in eqn.~\eqref{eq:Lagrangian_parallel}, we have
    \begin{equation}\label{eq:maximization_parallel1}
        \dfrac{\partial \hat{\xi}}{\partial \mathbf{T}^i}=-\dfrac{1}{(1+{\omega}^p)^2}\dfrac{\partial {\omega}^p}{\partial \mathbf{T}^i}\,\left(J\,\mathbf{T}\boldsymbol{:}\mathbf{D}\right)+\dfrac{1}{1+{\omega}^p}\,\dfrac{\partial \left(J\,\mathbf{T}\boldsymbol{:}\mathbf{D}\right)}{\partial \mathbf{T}^i}.
    \end{equation}
    Now let us look at the derivatives with respect to $\mathbf{T}^i$ separately. From the definition of $\omega^p$, its dependence on $\mathbf{T}^i$ stems from $\omega^p=\norm{\mathbf{T}^e}/\norm{\mathbf{T}^i}$. Thus, a routine calculation (see App.~\ref{sec:evolution_series}, derivation of eqn.~\eqref{eq:omega_Li}) yields
    \begin{equation}\label{eq:omega_Ti}
        \dfrac{\partial{\omega}^p}{\partial \mathbf{T}^i}=-{\omega}^p\,\dfrac{\mathbf{T}^i}{\norm{\mathbf{T}^i}^2}.
    \end{equation}
    The derivative of the total stress power with respect to $\mathbf{T}^i$ is rather tricky since the relation between $\mathbf{T}$ and $\mathbf{T}^i$ are not known unlike the rate of deformation tensors in eqn.~\eqref{eq:vel_grad_rel}. The only known relation between $\mathbf{T}$ and $\mathbf{T}^i$ are through their resultant stress power in eqn.~\eqref{eq:work_decomposition}. A direct derivative of eqn.~\eqref{eq:work_decomposition} results in
    \begin{equation}\label{eq:work_derivative_Ti}
        \dfrac{\partial \left(J\,\mathbf{T}\boldsymbol{:}\mathbf{D}\right)}{\partial \mathbf{T}^i}=J^i\,\mathbf{D}^i.
    \end{equation}
    Substituting eqns.~\eqref{eq:omega_Li} and \eqref{eq:work_derivative_Ti} into eqn.~\eqref{eq:maximization_parallel1}, we obtain
    \begin{equation}\label{eq:maximization_parallel_final}
        \mathbf{D}^i=\dfrac{1}{J^i}\,\left[\dfrac{\mathbf{T}^i}{\norm{\mathbf{T}^i}^2}\right]\,\dot{W}_2=\dfrac{1}{J^i}\,\left[\dfrac{\mathbf{T}^i}{\norm{\mathbf{T}^i}^2}\right]\,\xi.
    \end{equation}
    Eqn.~\eqref{eq:maximization_parallel_final} thus provides an explicit expression for $\mathbf{D}^i$ in terms of the driving force function, $\mathbf{T}^i$. One can easily verify that a contraction of eqn.~\eqref{eq:maximization_parallel_final} with $\mathbf{T}^i$ yields the rate of dissipation function.

    In case of a series connection, the rate of deformation tensor $\mathbf{D}^i$  was allowed to vary keeping the driving force fixed. The resultant driving force $\mathbf{T}^i$ was then used to derive a relation between $\mathbf{T}$ and $\mathbf{T}^e$ as shown in eqn.~\eqref{eq:stress_relations_series}. For a parallel connection, the calculations have been carried out in a driving force space, i.e., the driving force $\mathbf{T}^i$ was allowed to vary while $\mathbf{D}^i$ remains fixed. Hence, in this case, one might expect to obtain a relationship between the rate of deformation tensors using the resultant expression for $\mathbf{D}^i$ in eqn.~\eqref{eq:maximization_parallel_final}. The multiplicative decomposition of the deformation gradient, however, already establishes a relationship between $\mathbf{D}$ and $\mathbf{D}^i$ and hence, it does not need to be explored further. 

    From the analysis so far, we have obtained expressions for $\mathbf{D}^i$ and $\mathbf{T}^i$ in terms of their thermodynamic conjugates in case of a series and a parallel connection respectively. But for a given problem, only the total deformation gradient and the Cauchy stress is known \emph{a priori}. In order to find a complete solution, one must obtain their individual part, i.e., either $\mathbf{D}^e$ and $\mathbf{D}^i$ or, $\mathbf{T}^e$ and $\mathbf{T}^i$. Since different constitutive relations need to be assumed to describe the behavior of different rheological elements, the expressions for these individual parts depend on the rheological elements under consideration. Hence, the individual kinematic or kinetic quantities can only be obtained once a specific material behavior is to be modeled.
    
    \section{Illustrations}\label{sec:illustrations}
    
    In this section, we demonstrate the utility of our proposed theory by developing constitutive models for three different rheological networks associated with a standard linear solid (SLS), elastic-perfectly plastic (EPP) and an elastic-plastic solid exhibiting strain hardening behavior (EPH). The 1-D network of these models are shown in Fig.~\ref{fig:Sls_EPPM_EPSH}. The notations for the kinematic and kinetic quantities and, the expended stress power used in the different material models are listed in Table~\ref{table:Notation_table}.

   \begin{figure}[tbp]
	\centering
	\begin{subfigure}{0.35\linewidth}
		\includegraphics[width=\linewidth]{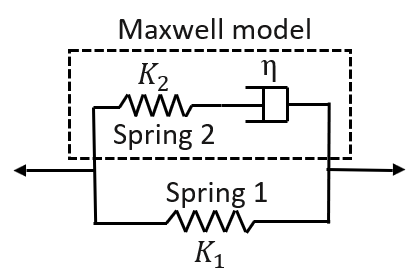}
		\caption{}
		\label{fig:SLS}
	\end{subfigure}
	\begin{subfigure}{0.35\linewidth}
		\includegraphics[width=\linewidth]{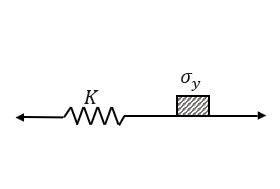}
		\caption{}
		\label{fig:EPPM}
	\end{subfigure}
	\begin{subfigure}{0.4\linewidth}
	        \includegraphics[width=\linewidth]{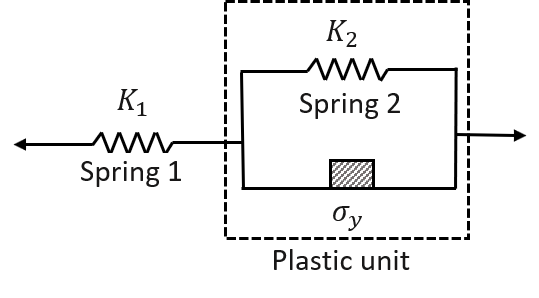}
	        \caption{}
	        \label{fig:EPSH}
         \end{subfigure}
	\caption{The rheological network of: (a) standard linear solid model (b) Elastic-perfectly plastic materials and (c) Elastoplastic materials with strain hardening.}
	\label{fig:Sls_EPPM_EPSH}
\end{figure}

    \subsection{Standard linear solid model}

    A standard linear solid (SLS) model is widely used in modeling viscoelastic behavior of materials within a small-strain theory. This model consists of a linear spring in a parallel connection with a Maxwell model i.e.,  a network of another linear spring and a dashpot joined in a series connection as shown in Fig.~\ref{fig:SLS}. For the finite deformation version of this rheological network, we first need to divide the network into its constituent parts. Let us first consider the first linear spring in a parallel connection with the Maxwell model. Note that the linear spring will be associated with a purely non-dissipative process whereas the Maxwell model, as a whole, acts like a dissipative system. We assume that the total applied power, Cauchy stress $\mathbf{T}$ and the rate of deformation tensor $\mathbf{D}$ for the entire network is known \emph{a priori}. We need to determine the individual components of the total power expended during the deformation of the spring and the Maxwell model. Since these two rheological elements are in a parallel connection, by Post.~\ref{Post:2}, the ratio of power in the spring and the Maxwell model is equal to the ratio of the norms of their respective Cauchy stress tensor, i.e., $\dot{\psi}_1/\xi=\norm{\mathbf{T}^e_1}/\norm{\mathbf{T}^i_M}$. Now by using Eqn.~\eqref{eq:maximization_parallel_final}, one can write the rate of deformation tensors behind the deformation of the Maxwell model as
    \begin{equation}\label{eq:Di_maxwell}
        \mathbf{D}^i_M=\dfrac{1}{J^i_M}\,\left[\dfrac{\mathbf{T}^i_M}{\norm{\mathbf{T}^i_M}}\,\left(\dfrac{\dot{W}}{\norm{\mathbf{T}^e_2}+\norm{\mathbf{T}^i_M}}\right)\right]
    \end{equation}
    where $J^i_M=J^e_2\,J^i_d$. Note that in eqn.~\eqref{eq:Di_maxwell}, the rate of deformation for the Maxwell model can be expressed completely in terms of the total power expended and the individual kinetic terms corresponding to the linear spring and the Maxwell model. The individual kinetic terms can be obtained once a specific form for the Helmholtz potential and the rate of dissipation function are assumed. 

    In view of eqn.~\eqref{eq:Di_maxwell}, now the relevant kinematic, kinetic and the stress power across the Maxwell model are known. Let us now derive the individual driving forces and rate of deformation tensors for the second linear spring and the dashpot that form the Maxwell model. Since these two rheological elements are in a series connection, the ratio of stress power is the same as the ratio of the norms of the individual rate of deformation tensors across these elements by Post.~\ref{Post:1}. As mentioned before, when the rate of dissipation function (or the Helmholtz potential) is assumed to have a specific form, the kinematic (or kinetic, depending on the connection) quantities can be obtained. For instance, if the rate of dissipation function for the dashpot is assumed to have the form 
    \begin{equation}
        \xi=\mathbf{D}^i_d\boldsymbol{:}\mathbb{C}_d\,\mathbf{D}^i_d
    \end{equation}
    where $\mathbb{C}_d$ is a fourth-order compliance tensor in the spirit of its similarity with a small-strain theory, one can obtain an implicit expression for the rate of deformation tensor of the dashpot in terms of the power expended in the Maxwell model as
    \begin{equation}\label{eq:Di_dashpot}
    \dfrac{1}{\norm{\mathbf{D}^i_d}}{\left(\mathbf{D}^i_d\boldsymbol{:}\mathbb{C}_d\mathbf{D}^i_d\right)\left(\norm{\mathbf{D}_M-\mathbf{D}^i_d}+\norm{\mathbf{D}^i_d}\right)}=\dot{W}_M.
    \end{equation} 
    Therefore, the only remaining task is to find the driving forces in terms of the stress power and these rate of deformation tensors. By employing eqn.~\eqref{eq:Ti_final}, we get the driving force for the dashpot as
    \begin{equation}\label{eq:Ti_dashpot}
        \mathbf{T}^i_d=\dfrac{1}{J^i_d\left(\norm{\mathbf{D}^e_2}+\norm{\mathbf{D}^i_d}\right)}\left[\dfrac{\norm{\mathbf{D}^e_2}}{\norm{\mathbf{D}^i_d}^2}\,\dot{W}\mathbf{D}^i_d+\norm{\mathbf{D}^i_d}\mathbf{F}^{e^T}_2\,\mathbf{T}_M\,\mathbf{F}^{e^{-T}}_2\right].
    \end{equation}
    The other driving forces and rate of deformation tensors can be found by using eqns.~\eqref{eq:DeDp} and \eqref{eq:work_decomposition}.

    From the solution of the SLS model (eqns.~\eqref{eq:Di_maxwell} and~\eqref{eq:Ti_dashpot}), it is imperative that not always a closed form analytical solution for the system can be obtained. In fact, these solutions depend on the assumed specific form for the Helmholtz potential and the rate of dissipation function. In this case, it is instructive to employ a suitable numerical technique to obtain specific solutions by using an iterative scheme. Moreover, one can observe that eqn.~\eqref{eq:Di_dashpot} is also an implicit equation in $\mathbf{D}^i_d$ and thus, needs to be solved numerically using an iterative method. The use of numerical techniques in finding solutions for dissipative systems is quite common, even within the limits of a small-strain theory~\cite{simo2006}. The development of such algorithms and finding specific semi-analytical solutions, however, is beyond the scope of the current work and will be the subject of future research. 

    \subsection{Elastic-perfectly plastic materials}

    Next we consider the case of an elastic-perfectly plastic (EPP) material. In a small-strain theory, this material is represented by a series connection between a linear spring and a frictional block. Although this idealized material is not of particular interest from a connection point of view, but it poses an interesting problem due to the constitutive behavior of the frictional block as shown in eqn.~\eqref{eq:spring_slider_series}.

    The fundamental premise of our developed model is based on a multiple natural configurations framework. In most traditional theories of plasticity, the existence of a yield surface is assumed which plays a key role in determining the response of a plastic material. On the other hand, in our theory, the rate of dissipation function takes the center stage~\cite{rajagopal2004}. Therefore, we consider that when the total stress power reaches a particular threshold, $\overline{\xi}$, the frictional block gets activated. Hence, the constitutive behavior of the elastic-perfectly plastic material can be written as
    \begin{align}
        \begin{cases}
            &\dot{\psi}=\dot{W}=J\,\mathbf{T}\boldsymbol{:}\mathbf{D}\;\text{with}\;\xi=\mathbf{D}^i=0\implies \mathbf{D}^e=\mathbf{D},\; \text{when}\; \mathbf{T}\boldsymbol{:}\mathbf{D}<\overline{\xi}\\
            &\xi=\dot{W}=J\,\mathbf{T}\boldsymbol{:}\mathbf{D}\;\text{with}\;\dot{\psi}=\mathbf{D}^e=0\implies \mathbf{D}=\mathbf{D}^i\;\text{otherwise}
        \end{cases}
    \end{align}
    Note that the Post.~\ref{Post:1} has been used to establish that $\mathbf{D}^i=0$ when $\xi=0$ and vice versa. 

    \subsection{Elastoplasticity with strain hardening}

    As shown in Fig.~\ref{fig:EPSH}, unlike the elastic-perfectly plastic materials, here a spring is connected to a frictional block in a parallel connection. Although the fundamental tenet of our frictional block model (i.e., it is activated when the available rate of dissipation reaches a threshold value) is still the same, the parallel connection between the spring and the frictional block leads to a rather significant result.

    Here we start our analysis with the series connection between the first spring and the plastic unit. Since the frictional block (and hence, the plastic unit) is activated only when the rate of dissipation function reaches a certain value of $\overline{\xi}$, the entire applied stress power is expended in the elastic process when $\dot{W}<\overline{\xi}$. Therefore, one can write
    \begin{equation}\label{eq:EPH_elastic}
        J\mathbf{T}\boldsymbol{:}\mathbf{D}=\dot{\psi}_1\,\text{and}\,\xi=0\;\text{when}\; J\mathbf{T}\boldsymbol{:}\mathbf{D}<\overline{\xi}
    \end{equation}
    Since this process is a non-dissipative one, the rate of deformation of the plastic unit $\mathbf{D}^i_p=0$ which in turn, implies that $\mathbf{D}^e_1=\mathbf{D}$. Using eqn.~\eqref{eq:EPH_elastic} and the condition $\mathbf{D}^e_1=\mathbf{D}$, one can easily arrive at
    \begin{equation}\label{eq:EPH_elastic_T}
        \mathbf{T}=\left(\dfrac{J^e_1}{J}\right)\,\mathbf{T}^e_1.
    \end{equation}

    Next we come to the second case where $J\,\mathbf{T}\boldsymbol{:}\mathbf{D}\ge \overline{\xi}$. In this case, the frictional block is activated and a certain amount of stress power, $\overline{\xi}$ is dissipated during its motion. Therefore, the rest of the stress power $\dot{W}_p-\overline{\xi}$ is used in the deformation of the second spring. The relation between the power expended in the plastic unit $\dot{W}_p$ and the total power can be easily obtained using Post.~\ref{Post:1} and the relationship~\eqref{eq:DeDp}. If one specifies a particular form for the Helmholtz potential function, then the relevant kinematic quantity associated with the second spring can be easily determined. For instance, if the Helmholtz potential is assumed to have the form 
    \begin{equation}\label{eq:helhmholtz_EPH}
        \psi=\dfrac{1}{2}\mathbf{E}_2\boldsymbol{:}\mathbf{C}_s\mathbf{E}_2,
    \end{equation}
    then one can easily obtain a differential equation for $\mathbf{E}_2$ as
    \begin{equation}\label{eq:EPH_E2}
        \mathbf{C}_s\mathbf{E}_2\boldsymbol{:}\dot{\mathbf{E}_2}=J_p\,\mathbf{T}_p\boldsymbol{:}\mathbf{D}_p-\overline{\xi}.
    \end{equation}
    Here $\mathbb{C}_s$ is a fourth-order elastic compliance tensor. Note that just like in the case of an SLS model, the eqn.~\eqref{eq:EPH_E2} needs to be solved by a numerical iterative method.

    Now let us understand the implication of the assumed yield condition in case of a strain-hardening elastic-plastic material. As mentioned earlier, in this case, the rate of dissipation function is restricted to a constant value of $\overline{\xi}$ and thus, one can write 
    \begin{equation}\label{eq:EPH_plastic}
    J^i\,\mathbf{T}^i\boldsymbol{:}\mathbf{D}^i=\overline{\xi}.
    \end{equation}
    Let us first assume that the rate of deformation tensor as well as the plastic driving force are allowed to vary for this model. Since the kinematic and kinetic quantities must satisfy eqn.~\eqref{eq:EPH_plastic}, they need to vary commensurately to keep the total rate of dissipation constant, i.e., if $\mathbf{T}^i$ increases, then $\mathbf{D}^i$ should proportionately decrease and vice versa. However, in view of eqn.~\eqref{eq:maximization_parallel_final}, $\mathbf{D}^i$ must be proportional to the driving force $\mathbf{T}^i$. Therefore, both $\mathbf{T}^i$ and $\mathbf{D}^i$ must have a constant value for this material model. The constant value for $\mathbf{T}^i$ is equivalent to the yield condition used in 1-D frictional block element where the block is activated when the applied stress reaches a certain value, namely the yield stress. Note that this condition is a direct consequence of the parallel connection between a spring and a frictional block (eqn.~\eqref{eq:maximization_parallel_final}) and has little significance in case of a series connection, i.e., for a elastic-perfectly plastic material.   

    \section{Summary}
    
    This work is aimed at extending the concept behind rheological connections, typically used in 1-D material model within a small-strain setting, to finite deformation. In the small strain setting, the rheological models are often described by the total stress or strain (or strain rate) distribution between the elements whereas their thrmodynamic conjugates remain the same. We observe that this is a mere consequence of the equilibrium and compatibility conditions. However, the more fundamental tenet of this type of model lies in the distribution of the applied stress power expended during the associated thermodynamic processes. The total stress power is distributed according to the ratio of strain rates or stresses across the elements in case of a series and a parallel connection respectively. In the case of finite deformation, these applied stress powers are expanded according to the ratio of norms of the rate of deformation or driving force tensors for a series and parallel connection, respectively. This work sets up the stage for extending specific material models that are described by rheological networks in a small-strain theory such as, a standard linear solid or an elastic-plastic material to a finite deformation framework. The proposed framework, however, depends on the assumed specific form of the Helmholtz potential and the rate of dissipation function. This framework leads to an implicit equation for the respective variables and therefore, a suitable numerical technique needs to be employed to obtain specific solutions using an iterative scheme. However, developing algorithms and finding specific semi-analytical solutions is beyond the scope of the current work and will be the subject of our future research. 
\bibliographystyle{acm}
\bibliography{connections_FD.bib}

\begin{thebibliography}{10}

\bibitem{acharya2000}
{\sc Acharya, A., and Bassani, J.~L.}
\newblock Lattice incompatibility and a gradient theory of crystal plasticity.
\newblock {\em Journal of the Mechanics and Physics of Solids 48}, 8 (2000),
  1565--1595.

\bibitem{altenbach2023}
{\sc Altenbach, H.}
\newblock Rheological modeling—historical remarks and actual trends in solid
  mechanics.
\newblock In {\em Advances in Mechanics of Time-Dependent Materials}. Springer,
  2023, pp.~1--16.

\bibitem{bingham1929}
{\sc Bingham, E.~C.}
\newblock Rheology. i. the nature of fluid flow.
\newblock {\em Journal of chemical education 6}, 6 (1929), 1113.

\bibitem{bingham1933}
{\sc Bingham, E.~C.}
\newblock The new science of rheology.
\newblock {\em Review of Scientific Instruments 4}, 9 (1933), 473--476.

\bibitem{blume1989}
{\sc Blume, J.~A.}
\newblock Compatibility conditions for a left cauchy-green strain field.
\newblock {\em Journal of elasticity 21}, 3 (1989), 271--308.

\bibitem{bouras2018}
{\sc Bouras, Y., Zorica, D., Atanackovi{\'c}, T.~M., and Vrcelj, Z.}
\newblock A non-linear thermo-viscoelastic rheological model based on
  fractional derivatives for high temperature creep in concrete.
\newblock {\em Applied Mathematical Modelling 55\/} (2018), 551--568.

\bibitem{brocker2012}
{\sc Br{\"o}cker, C., and Matzenmiller, A.}
\newblock Thermoviscoplasticity deduced from enhanced rheological models.
\newblock {\em PAMM 12}, 1 (2012), 327--328.

\bibitem{cermelli2001}
{\sc Cermelli, P., Fried, E., and Sellers, S.}
\newblock Configurational stress, yield and flow in rate--independent
  plasticity.
\newblock {\em Proceedings of the Royal Society of London. Series A:
  Mathematical, Physical and Engineering Sciences 457}, 2010 (2001),
  1447--1467.

\bibitem{gurtin2001}
{\sc Cermelli, P., and Gurtin, M.~E.}
\newblock On the characterization of geometrically necessary dislocations in
  finite plasticity.
\newblock {\em Journal of the Mechanics and Physics of Solids 49}, 7 (2001),
  1539--1568.

\bibitem{christensen2012}
{\sc Christensen, R.}
\newblock {\em Theory of viscoelasticity: an introduction}.
\newblock Elsevier, 2012.

\bibitem{cleja2000}
{\sc Cleja-Tigoiu, S., and Maugin, G.}
\newblock Eshelby's stress tensors in finite elastoplasticity.
\newblock {\em Acta Mechanica 139}, 1 (2000), 231--249.

\bibitem{epsteinmaugin1990}
{\sc Epstein, M., and Maugin, G.}
\newblock The energy-momentum tensor and material uniformity in finite
  elasticity.
\newblock {\em Acta Mechanica 83}, 3-4 (1990), 127--133.

\bibitem{epstein1996}
{\sc Epstein, M., and Maugin, G.~A.}
\newblock On the geometrical material structure of anelasticity.
\newblock {\em Acta Mechanica 115}, 1-4 (1996), 119--131.

\bibitem{eshelby1956}
{\sc Eshelby, J.}
\newblock The continuum theory of lattice defects.
\newblock {\em Solid state physics 3\/} (1956), 79--144.

\bibitem{eshelby1975}
{\sc Eshelby, J.}
\newblock The elastic energy-momentum tensor.
\newblock {\em Journal of elasticity 5}, 3-4 (1975), 321--335.

\bibitem{eshelby1951}
{\sc Eshelby, J.~D.}
\newblock The force on an elastic singularity.
\newblock {\em Philosophical Transactions of the Royal Society of London.
  Series A, Mathematical and Physical Sciences 244}, 877 (1951), 87--112.

\bibitem{ghobadi2021}
{\sc Ghobadi, E., Shutov, A., and Steeb, H.}
\newblock Parameter identification and validation of shape-memory polymers
  within the framework of finite strain viscoelasticity.
\newblock {\em Materials 14}, 8 (2021), 2049.

\bibitem{glass2008}
{\sc Glass, D.~H., Roberts, C.~J., Litsky, A.~S., and Weber, P.~A.}
\newblock A viscoelastic biomechanical model of the cornea describing the
  effect of viscosity and elasticity on hysteresis.
\newblock {\em Investigative ophthalmology \& visual science 49}, 9 (2008),
  3919--3926.

\bibitem{gupta2007}
{\sc Gupta, A., Steigmann, D.~J., and St{\"o}lken, J.~S.}
\newblock On the evolution of plasticity and incompatibility.
\newblock {\em Mathematics and Mechanics of Solids 12}, 6 (2007), 583--610.

\bibitem{he2020}
{\sc He, D., and Hu, Y.}
\newblock Nonlinear visco-poroelasticity of gels with different rheological
  parts.
\newblock {\em Journal of Applied Mechanics 87}, 7 (2020), 071010.

\bibitem{huber2000}
{\sc Huber, N., and Tsakmakis, C.}
\newblock Finite deformation viscoelasticity laws.
\newblock {\em Mechanics of materials 32}, 1 (2000), 1--18.

\bibitem{johnsen2019}
{\sc Johnsen, J., Clausen, A.~H., Grytten, F., Benallal, A., and Hopperstad,
  O.~S.}
\newblock A thermo-elasto-viscoplastic constitutive model for polymers.
\newblock {\em Journal of the Mechanics and Physics of Solids 124\/} (2019),
  681--701.

\bibitem{kiessling2015}
{\sc Kie{\ss}ling, R., Landgraf, R., and Ihlemann, J.}
\newblock Direct connection of rheological elements at large strains:
  Application to multiplicative viscoplasticity.
\newblock {\em PAMM 15}, 1 (2015), 313--314.

\bibitem{kiessling2016}
{\sc Kie{\ss}ling, R., Landgraf, R., Scherzer, R., and Ihlemann, J.}
\newblock Introducing the concept of directly connected rheological elements by
  reviewing rheological models at large strains.
\newblock {\em International Journal of Solids and Structures 97\/} (2016),
  650--667.

\bibitem{lakes2009}
{\sc Lakes, R.~S.}
\newblock {\em Viscoelastic materials}.
\newblock Cambridge university press, 2009.

\bibitem{li2017}
{\sc Li, Y., He, Y., and Liu, Z.}
\newblock A viscoelastic constitutive model for shape memory polymers based on
  multiplicative decompositions of the deformation gradient.
\newblock {\em International Journal of Plasticity 91\/} (2017), 300--317.

\bibitem{lion1997}
{\sc Lion, A.}
\newblock A physically based method to represent the thermo-mechanical
  behaviour of elastomers.
\newblock {\em Acta Mechanica 123}, 1-4 (1997), 1--25.

\bibitem{lion2000}
{\sc Lion, A.}
\newblock Constitutive modelling in finite thermoviscoplasticity: a physical
  approach based on nonlinear rheological models.
\newblock {\em International Journal of Plasticity 16}, 5 (2000), 469--494.

\bibitem{paulFreed2020gnd}
{\sc Paul, S., and Freed, A.~D.}
\newblock Characterization of the geometrically necessary dislocations using a
  {G}ram--{S}chmidt factorization of the deformation gradient.
\newblock {\em Zeitschrift f\"ur angewandte Mathematik und Physik 71}, 6
  (2020), 196.

\bibitem{peleg1983}
{\sc Peleg, K.}
\newblock A rheological model of nonlinear viscoplastic solids.
\newblock {\em Journal of Rheology 27}, 5 (1983), 411--431.

\bibitem{rajagopal1998ii}
{\sc Rajagopal, K., and Srinivasa, A.}
\newblock Mechanics of the inelastic behavior of materials. part ii: Inelastic
  response.
\newblock {\em International Journal of Plasticity 14}, 10-11 (1998), 969--995.

\bibitem{rajagopal1998i}
{\sc Rajagopal, K., and Srinivasa, A.}
\newblock Mechanics of the inelastic behavior of materials—part 1,
  theoretical underpinnings.
\newblock {\em International Journal of Plasticity 14}, 10-11 (1998), 945--967.

\bibitem{rajagopal2004}
{\sc Rajagopal, K.~R., and Srinivasa, A.~R.}
\newblock On thermomechanical restrictions of continua.
\newblock {\em Proceedings of the Royal Society of London. Series A:
  Mathematical, Physical and Engineering Sciences 460}, 2042 (2004), 631--651.

\bibitem{rajagopal2005}
{\sc Rajagopal, K.~R., and Srinivasa, A.~R.}
\newblock On the role of the eshelby energy-momentum tensor in materials with
  multiple natural configurations.
\newblock {\em Mathematics and mechanics of solids 10}, 1 (2005), 3--24.

\bibitem{reiner1929}
{\sc Reiner, M.}
\newblock The general law of flow of matter.
\newblock {\em Journal of Rheology 1}, 1 (1929), 11--20.

\bibitem{reisinger2020}
{\sc Reisinger, A.~G., Frank, M., Thurner, P.~J., and Pahr, D.~H.}
\newblock A two-layer elasto-visco-plastic rheological model for the material
  parameter identification of bone tissue.
\newblock {\em Biomechanics and Modeling in Mechanobiology 19\/} (2020),
  2149--2162.

\bibitem{shutov2008}
{\sc Shutov, A., and Krei{\ss}ig, R.}
\newblock Finite strain viscoplasticity with nonlinear kinematic hardening:
  Phenomenological modeling and time integration.
\newblock {\em Computer Methods in Applied Mechanics and Engineering 197},
  21-24 (2008), 2015--2029.

\bibitem{shutov2017}
{\sc Shutov, A.~V., Larichkin, A.~Y., and Shutov, V.~A.}
\newblock Modelling of cyclic creep in the finite strain range using a nested
  split of the deformation gradient.
\newblock {\em ZAMM-Journal of Applied Mathematics and Mechanics/Zeitschrift
  f{\"u}r Angewandte Mathematik und Mechanik 97}, 9 (2017), 1083--1099.

\bibitem{simo2006}
{\sc Simo, J.~C., and Hughes, T.~J.}
\newblock {\em Computational inelasticity}, vol.~7.
\newblock Springer Science \& Business Media, 2006.

\bibitem{wong2011}
{\sc Wong, Y.-S., Stachurski, Z., and Venkatraman, S.}
\newblock Modeling shape memory effect in uncrosslinked amorphous biodegradable
  polymer.
\newblock {\em Polymer 52}, 3 (2011), 874--880.

\end{thebibliography}

\setcounter{section}{0}
	\renewcommand{\thesection}{\Alph{section}}
	\section{Derivation of the evolution equation for a series connection of rheological elements in finite deformation}\label{sec:evolution_series}
	\setcounter{equation}{0}
	\renewcommand{\theequation}{A.\arabic{equation}}

 In this section, we show the detailed calculation for \S~\ref{sec:finite_def_series}. In order to derive the evolution equation, we first need to maximize the rate of dissipation function with respect to the rate of deformation tensor, $\mathbf{D}^i$ by using a Lagrange multiplier method, i.e., by maximizing the Lagrangian $\mathbb{L}$ (defined in eqn.~\eqref{eq:lagrangian_series}), as
 \begin{equation}\label{eq:xi_Li}
 \dfrac{\partial\mathbb{L}^s}{\partial  \mathbf{D}^i} = 0 \implies \dfrac{\partial\bar{\xi}}{\partial\mathbf{D}^i} = -\dfrac{\lambda}{(1+\lambda)}\,\dfrac{1}{(1+{\omega}^s)^2}\,\dfrac{\partial{\omega}^s}{\partial\mathbf{D}^i}\,\left(J\mathbf{T}\boldsymbol{:}\mathbf{D}\right)+\dfrac{\lambda}{(1+\lambda)}\,\dfrac{1}{(1+{\omega}^s)}\,\dfrac{\partial}{\partial\mathbf{D}^i}(J\mathbf{T}\,\boldsymbol{:}\mathbf{D})
 \end{equation}
Note that the ratio of stress power, $\omega^s$ and the total rate of deformation tensor $\mathbf{D}$ are both implicit functions of $\mathbf{D}^i$. Therefore, the derivative of these functions need to be evaluated first. 
For the ratio of stress power, $\omega^s$, we have
 \begin{equation}\label{eq:omega_Li}
    \,\dfrac{\partial{\omega}^s}{\partial\mathbf{D}^i}=\,-\,\dfrac{\norm{\mathbf{D}^e}}{\norm{\mathbf{D}^i}^2}\,\dfrac{\partial\norm{\mathbf{D}^i}}{\partial\mathbf{D}^i} = \,-\,\dfrac{{\omega}^s}{\norm{\mathbf{D}^i}}\,\dfrac{\mathbf{D}^i}{\norm{\mathbf{D}^i}} = -\dfrac{{\omega}^s\mathbf{D}^i}{\norm{\mathbf{D}^i}^2}
\end{equation}
Now for the derivative of the total stress power, we first simplify the equation by using eqn.~\eqref{eq:vel_grad_rel} and obtain
\begin{equation}\label{eq:jT_L_Li}
    \dfrac{\partial}{\partial\mathbf{D}^i}(J\mathbf{T}\,\boldsymbol{:}\mathbf{D}) =  \dfrac{\partial}{\partial\mathbf{D}^i}\left[J\mathbf{T}\,\boldsymbol{:}\mathbf{D}^e \, + \,J\mathbf{T}\,\boldsymbol{:}\mathbf{D}^i\right]= J\mathbf{T}
 \end{equation} 
 Substituting eqn.~\eqref{eq:omega_Li} and ~\eqref{eq:jT_L_Li} into eqn.~\eqref{eq:xi_Li}, we finally have the evolution equation as
\begin{equation}\label{eq:bar_xi_Li}
   \dfrac{\partial \bar{\xi}}{\partial \mathbf{D}^i}=\dfrac{\lambda}{1+\lambda}\left[\dfrac{1}{(1+{\omega}^s)^2}\,\dfrac{{\omega}^s\,\mathbf{D}^i}{\norm{\mathbf{D}^i}^2}\,\left(J\,\mathbf{T}\boldsymbol{:}\mathbf{D}\right)+\dfrac{J\mathbf{T}}{1+{\omega}^s}\right]  
\end{equation}
Note that the left hand side of eqn.~\eqref{eq:bar_xi_Li} is essentially the driving force $\mathbf{T}^i$. Now we use this equation to obtain more insights into the material models. Contracting eqn.~\eqref{eq:bar_xi_Li} with $\mathbf{D}^i$, we get
\begin{equation}\label{eq:contraction_1}
    J^i\,\mathbf{T}^i\boldsymbol{:}\mathbf{D}^i = \dfrac{\lambda}{1+\lambda}\dfrac{{\omega}^s}{(1+{\omega}^s)^2}\,\left(J\,\mathbf{T}\boldsymbol{:}\mathbf{D}\right)+\dfrac{\lambda}{1+\lambda}\,\dfrac{1}{\left(1+{\omega}^s\right)}\left(J\,\mathbf{T}\boldsymbol{:}\mathbf{D}^i\right)
\end{equation}
The Lagrange multiplier can be found by satisfaction of the constraint equation~\eqref{eq:xi_constraint_series} as $\dfrac{\lambda}{1+\lambda}=1$. Therefore, one is left with
\begin{equation}  
    J^i\,\mathbf{T}^i\boldsymbol{:}\mathbf{D}^i=\dfrac{\omega^s}{(1+{\omega}^s)^2}\,\left(J\,\mathbf{T}\boldsymbol{:}\mathbf{D}\right)+\dfrac{1}{(1+{\omega}^s)}\left(J\,\mathbf{T}\boldsymbol{:}\mathbf{D}^i\right)
\end{equation}
since $J^i\,\mathbf{T}^i\boldsymbol{:}\mathbf{D}^i=\xi$ and from eqn.~\eqref{eq:xi_constraint_series}
\begin{equation}\label{eq:W1_dot}
    \xi=\dfrac{\omega^s}{(1+{\omega}^s)}\,\xi+\dfrac{1}{(1+{\omega}^s)}\left(J\,\mathbf{T}\boldsymbol{:}\mathbf{D}^i\right) \implies (J^i\,\mathbf{T}^i-J\,\mathbf{T})\boldsymbol{:}\,\mathbf{D}^i=0
\end{equation} 
where $\mathbf{T}^i$ is the driving force behind the motion associated with $\mathbf{D}^i$. Note that eqn.~\eqref{eq:W1_dot} has the same form as the normality condition in a traditional plasticity theory within a small-strain framework, i.e., $(\boldsymbol{\sigma}-\boldsymbol{\sigma^*})\cdot \boldsymbol{\dot{\epsilon}}^p=0$ where $\boldsymbol{\sigma}$ and $\boldsymbol{\dot{\epsilon}}^p$ represent stress and plastic strain rate at a point on the yield surface and $\boldsymbol{\sigma^*}$ is any admissible stress within the yield surface. In this case, however, $\boldsymbol{T}^i$ has a specific physical significance and represents an inelastic driving force measured from the relevant natural configuration. In other words, $\boldsymbol{T}^i  $ is \emph{not} just any admissible stress on the yield surface. Moreover, the theory under discussion is valid for any rheological network irrespective of their physical significance. Hence, although the structure of the equation seems similar to that of a traditional plasticity framework, one must not confuse the developed theory with the physical understanding of a small-strain plasticity. Here the term $(J^i\,\mathbf{T}^i-J\,\mathbf{T})$ does not need to be normal to $\boldsymbol{D}^i$. Since eqn.~\eqref{eq:W1_dot} is valid for any arbitrary $\mathbf{D}^i$, we obtain
\begin{equation}
    J^i\,\mathbf{T}^i=J\,\mathbf{T}.
\end{equation}

\end{document}